%% file: Main_Text2.tex
\documentclass[11pt,a4paper,twoside]{article}
\usepackage[utf8]{inputenc}
\usepackage{cite}

\usepackage{amsmath}

\usepackage{amssymb}

\usepackage[english]{babel}

\usepackage{epsfig}

\usepackage{a4wide}
\usepackage{dsfont}
\usepackage{graphics, color}
\def\rhs{r.h.s.\ }
\newcommand{\la}{\lambda}
\newcommand{\om}{\omega}
\DeclareMathOperator{\sh}{sh}
\DeclareMathOperator{\ch}{ch}

\title{Finite temperature correlation functions from discrete functional equations}

\author{Britta Aufgebauer \ and Andreas Kl\"umper \\
 \parbox{0.9 \linewidth}{\vspace{0.4 \baselineskip}\centering
  Fachbereich C -- Physik, Bergische Universit\"at Wuppertal, \\ 42097
    Wuppertal, Germany}}
\date{\today}
\begin{document}
\maketitle
\begin{abstract}
We present a new approach to the static finite temperature correlation
functions of the Heisenberg chain based on functional equations. An
inhomogeneous generalization of the $n$-site density operator is
considered. The lattice path integral formulation with a finite but arbitrary
Trotter number allows to derive a set of discrete functional equations
with respect to the spectral parameters. We show that these equations yield a
unique characterisation of the density operator. Our functional equations are a discrete version of the reduced q-Knizhnik-Za\-mo\-lod\-chi\-kov equations which played a central role in the
study of the zero temperature case. As a natural result, and independent of
the arguments given by Jimbo, Miwa, and Smirnov (2009) we prove that the inhomogeneous finite temperature correlation functions have the same remarkable structure as for zero temperature: they are a sum of products of nearest-neighbor correlators.
\end{abstract}

\section{Introduction}
The seminal model of integrable lattice systems is the spin-1/2 Heisenberg
chain whose spectrum can be exactly calculated by Bethe ansatz \cite{Bethe31,Babook}. Many
properties of the model have been calculated, e.g. the thermodynamical
potential \cite{T71, G71} and the scaling dimensions \cite{BCN86, C86, VW85, W87, K88, KB90, KWZ93} describing the long distance asymptotics of all correlation functions.

The calculation of space and time dependent correlation functions is a
notorious problem. A well-known procedure is the form factor approach which is
based on a spectral decomposition of the space and time dependent two-point
function by use of the eigenvalues of the Hamiltonian and of the matrix
elements of the local operators with respect to the eigenstates of the
Hamiltonian. For interacting systems the calculation of matrix elements is nontrivial \cite{Smirnov92, JM95, KMT99a}. In applications the spectral sum is carried out over sectors with restricted particle number \cite{BKM98,KMBFM97,AKMW06,SST04,CMP08,KMCB11}. Recently, analytic summations over infinite classes of states were achieved \cite{KMST05, KKMST09, KKMST11a, KKMST11b}.
 
Methodically different are approaches that aim at the computation of the reduced density
matrix. Based on for instance the representation theory of quantum algebras \cite{JMMN92, JM95}, functional equations \cite{JiMi96}, or the algebraic Bethe ansatz \cite{KMT99b, KMST04c, GKS04a, GKS05, BoGo09}, multiple integral
representations for the elements of the density matrix were derived. 

In a series of works \cite{BoKo01, BKS03, BKS04c, BJMST04a, BJMST04b} it was
shown that the general ground-state correlation functions of an inhomogeneous
version of the model can be represented by algebraic expressions involving
only a few independent functions describing nearest-neighbor correlations.

An exponential formula for the reduced density matrix was obtained in
\cite{BJMST05b,BJMST06}. On the basis of an explicit calculation for short distances of local operators, this formula was conjectured in \cite{BGKS06,DGHK07,BGKS07 } to be also valid for finite temperature or finite length. In \cite{BJMST06b, BJMST08a} a
fermionic structure on the space of local operators was identified from which
a generating function of all local correlation functions was obtained
\cite{JMS08} for a very general inhomogeneous vertex model including the
finite temperature and the finite length Heisenberg chain as special cases.
Even in this most general situation the inhomogeneous correlation functions depend on only two functions \cite{JMS08, BoGo09}. 

The properties of the Heisenberg chain at finite temperature are obtained within
the quantum transfer matrix approach using the correspondence between the
Heisenberg chain and the six-vertex model. This method has been developed as an alternative approach \cite{B81, TS83, K87, SI87, SAW90, SNW90, T91, T91b, K92, K93, DV92, DV95}, see also the reviews \cite{KS03,K04}, to the thermodynamical Bethe ansatz which is based on combinatorics and the controversial string picture of the excited states of the Heisenberg chain. Here we consider independent spectral
parameters in both the chain and the quantum direction of the six-vertex model
to define an inhomogeneous $n$-site density operator as introduced in the work \cite{JMMN92}. For the zero
temperature case, a functional equation of Knizhnik-Zamolodchikov type \cite{KZ84, FrRe92} was
derived \cite{JM95, JiMi96}. At finite temperature the functional equation is not satisfied as it
receives finite temperature correction terms which seem unmanageable.

We find the functional equation approach to the computation of correlation
functions very important \cite{SABGKTT11}. Such an analysis is basically built only on the
fundamental properties of the $R$-matrix: the Yang-Baxter equation, the \lq
standard initial condition\rq , and also, if satisfied, the crossing symmetry.
Hence, a generalization of previous work to higher spin or higher rank models
looks most feasible along this way. Before doing so, it is important to first
understand the computation of the finite temperature correlations of the
Heisenberg chain by use of suitable functional equations.

Facing the above mentioned problem, the first and obvious intention is to
identify the finite temperature correction terms appearing in the functional
equations. In this paper, however, we perform an analysis based on a rather
contrary reasoning. We find that the correction terms vanish at certain
discrete points of the spectral parameters. Interestingly, the reduced
information is still sufficient to fix the full functional dependence of the
finite temperature correlations. We therefore call these equations \lq
discrete functional equations\rq. This paper is organized as follows. In section 2 we define the inhomogeneous
density matrix and derive the discrete functional equations in section 3 with the proof of uniqueness deferred to the appendix. The solution to the discrete functional equations for the $XXX$ case is presented in detail 
in section 4. Generalizations to the $XXZ$ case and for systems with an
$\alpha$-seam are found in the appendix.

\section{The inhomogeneous $n$-site density operator}
\subsection{The $R$-matrix}
The Hamiltonian of the spin-1/2 Heisenberg chain is given by
\begin{equation}
H_{XXX}=
\sum_{i=1}^{L}\left(\sigma_i^x\sigma_{i+1}^x+\sigma_i^y\sigma_{i+1}^y+ \Delta\sigma_i^z\sigma_{i+1}^z\right)\label{XXX-Hamiltonian}
\end{equation}
where $\sigma^x$, $\sigma^y$, and $\sigma^z$ are the Pauli matrices 
\begin{equation}
\sigma^x=\begin{pmatrix}0&1\\1&0
\end{pmatrix},\quad 
\sigma^y=\begin{pmatrix}0&-i\\i&0
\end{pmatrix},\quad
\sigma^z=\begin{pmatrix}1&0\\0&-1
\end{pmatrix}.\nonumber
\end{equation}
We denote the orthonormal eigenstates of $\sigma^z$ by $v^+$ (eigenvalue 1)
and $v^-$ (eigenvalue -1). We are interested in the thermodynamic limit of
(\ref{XXX-Hamiltonian}). Therefore, the boundary conditions are unimportant,
but we may impose periodic boundary conditions whenever arguments are based on
a finite size $L$.

The $R$-matrix of the classical counterpart of the Heisenberg chain, the
six-vertex model, is given by
\begin{equation}
\check{R}(\lambda_1, \lambda_2):=\begin{pmatrix}
a(\lambda_1, \lambda_2)&&&\\&1&b(\lambda_1, \lambda_2)&\\&b(\lambda_1, \lambda_2)&1&\\&&& a(\lambda_1, \lambda_2)
\end{pmatrix}.\nonumber
\end{equation}
For $\Delta=\ch(\eta)$ the functions $a$ and $b$ are defined as
\begin{equation}
a(\lambda_1, \lambda_2)=[1+\lambda_1-\lambda_2]_q,\quad b(\lambda_1, \lambda_2)=[\lambda_1-\lambda_2]_q.\label{XXZ_Funktionen}
\end{equation}
with
\begin{equation}
[x]_q:=\frac{\sh(\eta x)}{\sh(\eta)} \quad\mbox{for}\quad q=e^\eta.\nonumber
\end{equation}
In the limit $q\rightarrow 1$ the functions for the XXX case ($\Delta=1$) are obtained:
\begin{equation}
a(\lambda_1, \lambda_2)=1+\lambda_1-\lambda_2,\quad b(\lambda_1, \lambda_2)=\lambda_1-\lambda_2.
\end{equation}
The matrix elements of the $R$-matrix are considered as local Boltzmann
weights of the six-vertex model
\begin{equation}
\check{R}(\la_1, \la_2)(v^{\mu_1}\otimes v^{\mu_2})=\sum_{\sigma_1, \sigma_2} \check{R}(\la_1, \la_2)^{\mu_1 \mu_2}_{\sigma_1 \sigma_2}\; v^{\sigma_1}\otimes v^{\sigma_2}\nonumber
\end{equation}
and are graphically represented by vertices with spin variables at the ends of
the bonds and spectral parameters associated with the lines, see Fig.~\ref{fig:R}. The $R$-matrix enjoys the following properties:
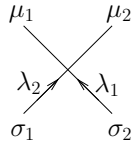
\begin{figure}[h!]
\centering
\resizebox{!}{2cm}{\input{R_ind}}
\caption{Graphical depiction of the matrix element $\check{R}(\la_1, \la_2)^{\mu_1 \mu_2}_{\sigma_1 \sigma_2}$.}
\label{fig:R}
\end{figure}

\noindent{\bf Yang-Baxter equation} (illustrated in Fig.~\ref{fig:YBE_UNI})
\begin{equation}
\check{R}_{1, 2}(\lambda_2,\lambda_3)\,\check{R}_{2, 3}(\lambda_1,\lambda_3)\,\check{R}_{1, 2}(\lambda_1,\lambda_2)
=\check{R}_{2, 3}(\lambda_1,\lambda_2)\,\check{R}_{1, 2}(\lambda_1,\lambda_3)\, \check{R}_{2, 3}(\lambda_2,\lambda_3),\label{YBEq}
\end{equation}
{\bf initial condition} (illustrated in Fig.~\ref{fig:ANF_CROSS_SING})
\begin{equation}
\check{R}(\lambda, \lambda)=  I\otimes I,\label{eq:Anf}
\end{equation}
and {\bf unitarity} (illustrated in Fig.~\ref{fig:YBE_UNI})
\begin{equation}
\check{R}(\lambda_2, \lambda_1)\,\check{R}(\lambda_1, \lambda_2)=C_U\; I\otimes I,\quad C_U:=a(\la_{1}-\la_2)a(-(\la_{1}-\la_2)). \label{eq:Uni}
\end{equation}
Furthermore it possesses the {\bf crossing symmetry} (illustrated in
Fig.~\ref{fig:ANF_CROSS_SING}).
For 
\begin{equation}
R(\lambda_1,\lambda_2):=P\, \check{R}(\lambda_1,\lambda_2)=\sum_{i,j=1}^4 \left(\varphi_i\otimes\psi_j\right)\nonumber
\end{equation}
with suitable maps $\varphi_i,\psi_j\in$ End$(V)$ and with $P$ denoting the permutation operator on two sites we find
\begin{equation}
\sum_{i,j} \left(\varphi_i\otimes\psi^t_j\right)=\left(S \otimes I\right)\,R(\lambda_2-1,\lambda_1)\,\left(S \otimes I\right), \label{Cross}
\end{equation}
where $S$ is defined by
\begin{equation}
S(v^+)=v^-;\quad S(v^-)=-v^+.\nonumber
\end{equation}
A graphical depiction of the matrix elements of $S$ with respect to the basis $v^+, v^-$ is introduced in Fig.~\ref{fig:S}.
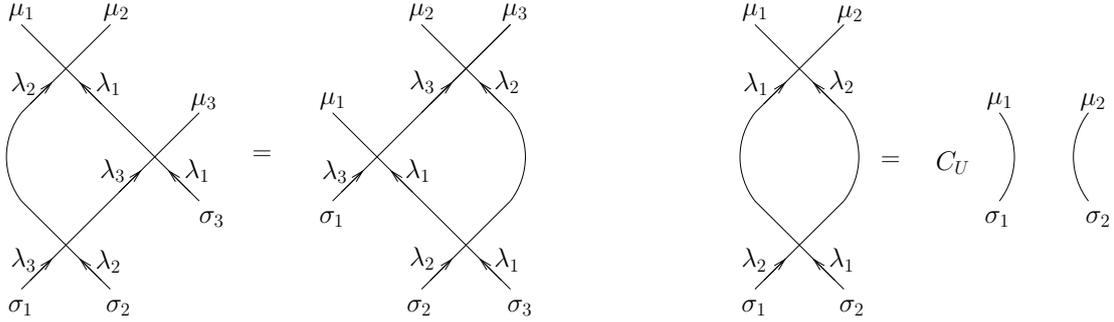
\begin{figure}[h!]
\centering
\resizebox{!}{5cm}{\input{YBE_UNI_ind_fig}}
\caption{Graphical depiction of the Yang-Baxter equation (left graph) and
  unitarity (right graph). A summation over all allowed values of the spin
  variables on closed bonds is understood implicitly. $\lambda_1, \lambda_2,
  \lambda_3$ denote spectral parameters assigned to lines; $\sigma_1,
  \sigma_2, \sigma_3$ and $\mu_1, \mu_2, \mu_3$ are spin variables assigned to open bonds. }
\label{fig:YBE_UNI}
\end{figure}
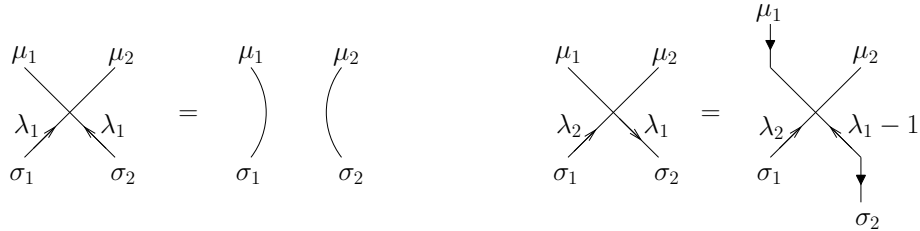
\begin{figure}[h!]
\centering
\resizebox{!}{3.2cm}{\input{ANF_CROSS_ind}}
\caption{Graphical depiction of initial condition (left graph) and crossing
  symmetry (right graph).}
\label{fig:ANF_CROSS_SING}
\end{figure}
\begin{figure}[h!]
\centering
\resizebox{!}{1.8 cm}{\input{ID_SING_alternativ}}
\caption{Matrix elements of the identity and the single site map $S$.}
\label{fig:S}
\end{figure}

\subsection{Construction of the inhomogeneous $n$-site density operator}\label{Construction Dn}
We study a semi-infinite six-vertex model on a regular lattice of height $N$
and infinite width, see Fig.~\ref{fig:Dn}. The width corresponds to the
length of the Heisenberg chain, the height corresponds to the Trotter number
of the imaginary time discretization. For a suitable choice of spectral
parameters assigned to the rows, the product of the $N$ row-to-row transfer
matrices approaches $\exp(-\beta H)$ in the limit $N\to\infty$. The following
analysis, however, is done for finite $N$ and the limit $N\to\infty$ is taken
at the very last stage.

A very important object for the study of the model on the semi-infinite
cylinder is the column-to-column transfer matrix. This matrix (in the
limit $N\to\infty$) is also known as the quantum transfer matrix of the
Heisenberg chain. We define the corresponding inhomogeneous quantum monodromy
matrix as 
\begin{equation}
\mathcal{T}(\la):=\check{R}_{N, N+1}(\la, \nu_N) \cdots \check{R}_{2, 3}(\la, \nu_2)\check{R}_{1,2}(\la, \nu_1).
\nonumber
\end{equation}
With respect to the basis $v^+, v^-$ in the space $V_\la$ the matrix is denoted as: 
\begin{equation}
\begin{pmatrix}
\mathcal{T}^+_+(\la)&\mathcal{T}^-_+(\la)\\\mathcal{T}^+_-(\la)&\mathcal{T}^-_-(\la)
\end{pmatrix}
,\quad \mathcal{T}^i_j(\la)\in \mbox{End}\left(V_Q\right),\quad V_Q:=V_{\nu_1}\otimes \cdots \otimes V_{\nu_N}\nonumber
\end{equation}
The quantum transfer matrix is obtained by taking the trace over the auxiliary space $V_\la$:
\begin{equation}
T(\la):=\mbox{tr}_{V_\la}\left(\mathcal{T}(\la)\right)=\mathcal{T}^+_+(\la)+\mathcal{T}^-_-(\la)\nonumber
\end{equation}
We define the inhomogeneous $n$-site density operator 
\begin{equation}
D_n(\la_1, \dots, \la_n)\in \mbox{End}(V_{\la_1}\otimes \cdots \otimes V_{\la_n})\nonumber
\end{equation}
in the following way:
\begin{equation}
D_n(\la_1, \dots, \la_n):=\lim_{k\rightarrow\infty}\frac{\mbox{tr}_{V_Q}\left([T(0)]^k \mathcal{T}(\la_1)\mathcal{T}(\la_2) \cdots  \mathcal{T}(\la_n) [T(0)]^k\right)}{\mbox{tr}_{V_Q}\left([T(0)]^k {T}(\la_1){T}(\la_2)\cdots {T}(\la_n) [T(0)]^k\right)}\label{D_Prae}
\end{equation}
\begin{figure}[h]
\centering
\resizebox{!}{6cm}{\input{Zahler_ind}}
\caption{Graphical depiction of
$\mbox{tr}_{V_Q}\left([T(0)]^k\mathcal{T}^{\mu_1}_{\sigma_1}(\la_1)\mathcal{T}^{\mu_2}_{\sigma_2}(\la_2)\cdots\mathcal{T}^{\mu_n}_{\sigma_n}(\la_n)[T(0)]^k\right)$}
\label{fig:Dn}
\end{figure}
Due to the commutativity of the quantum transfer matrices, $\left[T(\la),
  T(\mu)\right]=0$, the eigenvectors of $T(\la)$ can be chosen independently
of the spectral parameter $\la$. We assume that the spectral parameters
$\nu_1, \dots, \nu_N$ are chosen in such a way, that the leading eigenvalue of
$T(0)$ is non-degenerate. In this case only the (normalised) eigenvector $\Phi_0$,
corresponding to the leading eigenvalue $\Lambda_0$, contributes to the trace:
\begin{equation}
D_n(\la_1, \dots, \la_n)=\frac{\left<\Phi_0\right|\mathcal{T}(\la_1)\mathcal{T}(\la_2)\cdots \mathcal{T}(\la_n)\left|\Phi_0\right>}{\prod_{i=1}^n \Lambda_0(\la_i)}.\label{D_Fin}
\end{equation}
Note that for the physical values stated in equation (\ref{phys_nus}) there is
indeed a finite gap between the leading eigenvalue and the rest of the spectrum even
in the Trotter limit $N\rightarrow \infty$. For a proof see \cite{SI87, SAW90} and chapter 13 of the book \cite{EFGKK05} where an intuitive argument based on the high temperature case is given.

The concatenation of transfer respectively monodromy matrices in formulae
(\ref{D_Prae}), (\ref{D_Fin}) is to be understood with respect to the quantum
space $V_{\nu_1}\otimes \cdots \otimes V_{\nu_N}$, meaning that the matrix
elements are given by
\begin{equation}
{D}^{\mu_1 \dots \mu_n}_{\sigma_1 \dots \sigma_n}(\la_1, \dots, \la_n)=\frac{\left<\Phi_0\right|\mathcal{T}^{\mu_1}_{\sigma_1}(\la_1)\mathcal{T}^{\mu_2}_{\sigma_2}(\la_2)\cdots \mathcal{T}^{\mu_n}_{\sigma_n}(\la_n)\left|\Phi_0\right>}{\prod_{i=1}^n \Lambda_0(\la_i)}.\label{Mat_el}
\end{equation}
The $n$-site density operator for the infinite Heisenberg chain in a heat bath with temperature $T$ is obtained from $D_n(\la_1, \dots, \la_n)$ by first specialising the parameters $\nu_j, 1 \le j \le N$ in the quantum space (for even Trotter number) to the values
\begin{equation}
\nu_{2k+1}=\frac{\beta}{N}, \quad \nu_{2k}=-\frac{\beta}{N}+1\quad k=1,\dots, \frac{N}{2},\label{phys_nus}
\end{equation}
then performing the Trotter limit $N\rightarrow \infty$ and taking the
so-called homogeneous limit $\la_i\rightarrow 0$ for $1\le i \le n$.

\section{The discrete functional equations}
For zero temperature the density operator satisfies the functional equation
\begin{subequations}
\begin{equation}
A_n(\la_1, \dots, \la_n) \left[D_n(\la_1,\dots, \la_{n-1},\la_n)\right]=D_n(\la_1,\dots, \la_{n-1},\la_n-1)\label{AD}
\end{equation}
for arbitrary complex $\la_n$. Under suitable asymptotic conditions and analyticity the solution to this functional equation is unique \cite{BJMST04a, BJMST04b}. At finite temperature resp. finite Trotter number the functional equation does not hold for arbitrary $\la_n$. The proof of (\ref{AD}) for $T=0$ is based on so-called \lq Z-invariance\rq\,arguments \cite{Babook} relating the partition functions of different lattices. At $T=0$ certain terms disappear as they can be viewed as boundary terms at infinity. For finite Trotter number these terms matter. When we attempted to quantify these terms we were surprised to realize that the unwanted terms do not appear for the finite set of arguments
\begin{equation}
\la_n=\nu_1, \dots, \nu_N.\label{12b}
\end{equation}
\end{subequations}
Interestingly, equation (\ref{AD}) for (\ref{12b}) and asymptotic and analytic properties of the density operator are sufficient to uniquely fix the solution \cite{SABGKTT11} as will be shown at the end of this section.

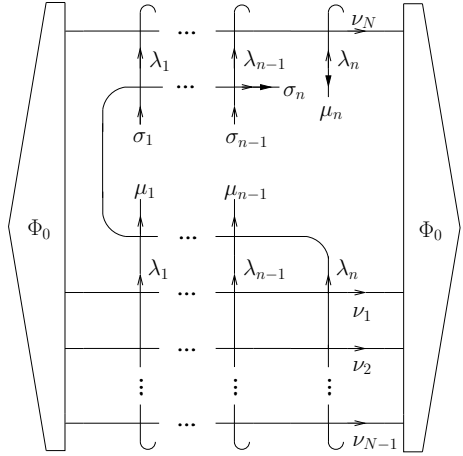
\begin{figure}[h]
\centering
\resizebox{!}{6 cm}{\input{Def_A}}
\caption{Graphical illustration of the matrix element
  $\left(A_n\left[D_n(\la_1, \dots, \la_n)\right]\right)^{\mu_1 \dots
    \mu_n}_{\sigma_1 \dots \sigma_n}$ which is basically given by the
  partition function of a lattice as shown in Fig.~\ref{fig:Dn} extended by a
  sequence of $R$-matrices. Here, the infinitely many columns in the
  thermodynamical limit have been replaced by the leading left and right
  eigenstates $\Phi_0$.  Note that for the normalization of the density matrix
  we would have to multiply it with $C_A^{-1} \,\prod_{i=1}^n \Lambda_0(\la_i)$.}
\label{fig:A_op}
\end{figure}   
\begin{figure}[h!]
\centering
\resizebox{!}{6cm}{\input{A_plus}}
\caption{Graphical illustration of $\left(A_n [D_{n+1}(\la_1,
  \dots, \la_n, \la_{n}-1)]\right)^{\mu_1 \dots \mu_n \mu_{n+1}}_{\sigma_1 \dots
    \sigma_n \sigma_{n+1}}$ multiplied by $C_A^{-1}\,\left[\prod_{i=1}^n
    \Lambda_0(\la_i)\right]\Lambda_0(\la_n-1)$. Note that a contraction of the
  `$\la_n$-line' will reduce the object to a $n$-site density matrix. The same
  holds true for a contraction of the `$\la_n-1$-line'.}
\label{fig:qKZ1}
\end{figure}
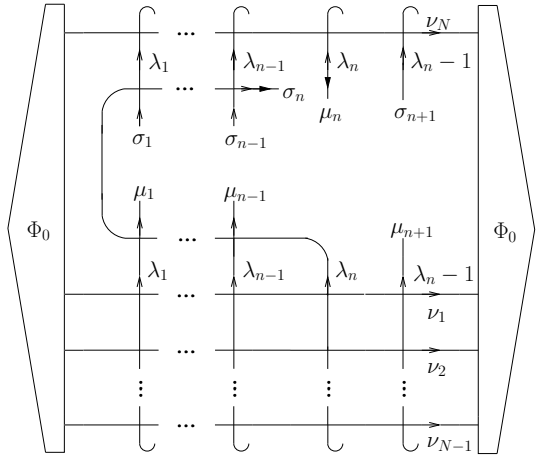
\begin{figure}[h!]
\centering
\resizebox{!}{6.2cm}{\input{A_plus2}}
\caption{LHS: Using the crossing symmetry for the quantum monodromy matrix
  $\mathcal{T}(\la_n-1)$ acting on site $n+1$ the direction of the
  $(\la_n-1)$-line is changed, thereby raising the spectral parameter by
  one. For the special case that $\la_n$ is equal to one of the parameters
  $\nu_i$ (we choose $\nu_1$) we find the RHS by use of the initial condition
  and unitarity. RHS: the two traces in eqs.~(\ref{tr1}) and (\ref{tr2}),
  contractions over the pair $(\sigma_n,\mu_n)$ and
  $(\sigma_{n+1},\mu_{n+1})$, respectively, yield the same object.}
\label{fig:qKZ}
\end{figure}
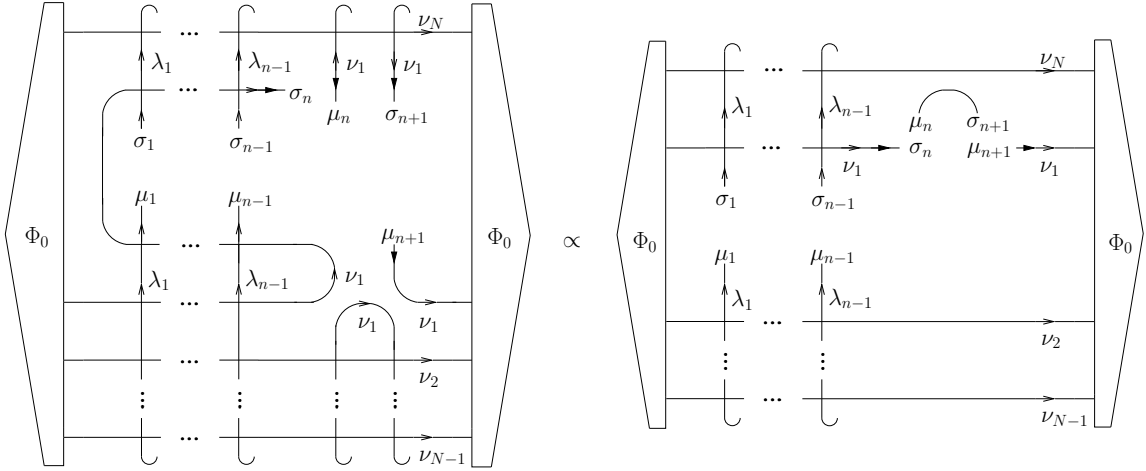

The operator 
\begin{equation}
A_n(\la_1, \dots, \la_n)\in \mbox{End}\left(\mbox{End}(V_{\la_1}\otimes\dots\otimes V_{\la_n})\right)\nonumber
\end{equation}
is constructed from $R$-matrices, see Fig.~\ref{fig:A_op}. The image of an
operator $B\in \mbox{End}(V^{\otimes n})$ is defined as\footnote{The
  definition of $A_n$ is more easily stated by using the duality
  $\mbox{End}(V)\cong V\otimes V^*$. (Compare eq. (\ref{A_2}).)}
\begin{multline}
A_n(\lambda_1, \dots,\lambda_n)[B]:= C_A\; \mbox{tr}_{V_n}\Bigl(\Bigr .\check{R}_{1,2}(\lambda_1, \lambda_n)\,\cdots\,\check{R}_{n-1,n}(\lambda_{n-1}, \lambda_n)\,(Pr_s)_{n, n+1}\\ \left(B\otimes I_{n+1}\right)\, \check{R}_{n-1,n}(\lambda_n, \lambda_{n-1})\,\cdots\,\check{R}_{2,3}(\lambda_n, \lambda_2)\,\check{R}_{1,2}(\lambda_n, \lambda_1)  \Bigl .\Bigr).\label{Def:An}
\end{multline}
The constant $C_A$ is defined as 
\begin{equation}
C_A:=\prod_{l=1}^{n-1}C_U^{-1}(\lambda_n, \lambda_l)\nonumber
\end{equation}
and $Pr_s:=\left|\right.s\left.\right>\left<\right.s\left.\right|$ denotes the (unnormalised) projector onto the 2-site $sl_2$-singlet
\begin{equation}
s= v^+\otimes v^- -  v^-\otimes v^+.\label{singlet}
\end{equation}
In order to derive the discrete functional equations (\ref{AD}) for (\ref{12b}) we consider the inhomogeneous density operator on $(n+1)$ sites with the specialization $\la_{n+1}=\la_n-1$:
\begin{equation}
D_{n+1}(\la_1, \dots, \la_n, \la_{n}-1)=\frac{\left<\Phi_0\right|\mathcal{T}(\la_1)\mathcal{T}(\la_2)\cdots\mathcal{T}(\la_n)\mathcal{T}(\la_n-1)\left|\Phi_0\right>}{\Lambda_0(\la_1)\Lambda_0(\la_2)\cdots \Lambda_0(\la_n)\Lambda_0(\la_n-1)}\nonumber
\end{equation}
We consider the action of $A_n$ on the first $n$ sites of $D_{n+1}(\la_1, \dots, \la_n, \la_{n}-1)$, see figure (\ref{fig:qKZ1}). Taking the trace of the image $A_n [D_{n+1}(\la_1, \dots, \la_n, \la_{n}-1)]$ with respect to the $n$-th respecively $(n+1)$-th space yields
\begin{align}
\mbox{tr}_{V_n}\left(A_n [D_{n+1}(\la_1, \dots, \la_n, \la_{n}-1)]\right)&=D_{n}(\la_1, \dots, \la_{n}-1) \label{tr1},\\
\mbox{tr}_{V_{n+1}}\left(A_n [D_{n+1}(\la_1, \dots, \la_n, \la_{n}-1)]\right)&=A_n[D_{n}(\la_1, \dots, \la_{n})]\label{tr2}
\end{align}
for all values of $\la_n$. For the special case that $\la_n$ is equal to one
of the horizontal inhomogenieties $\nu_1,\dots, \nu_N$, an important
simplification occurs due to the standard initial condition (equation (\ref{eq:Anf})) leading to the
factorization of two $R$-matrices. Consequently the left hand sides of equations
(\ref{tr1}) and (\ref{tr2}) become equal and hence the right hand sides are identical. In other words we obtain equation
(\ref{AD}). The graphical proof is given in Fig.~\ref{fig:qKZ}. 
\footnote{It is possible to state equation (\ref{AD}) for a more general model, including a magnetic field and disorder parameter. This issue is discussed in appendix \ref{Funk_alpha}.}

Note that the presented construction can be substantially generalized to other boundary conditions. If $\left<\Phi_0\right|$ and $\left|\Phi_0\right>$ in (\ref{D_Fin}) and Fig.~\ref{fig:A_op} etc.~are replaced by arbitrary left and right eigenstates $\left<\Phi_l\right|$ and $\left|\Phi_r\right>$ (to the same or different eigenvalues of the quantum transfer matrix) all arguments leading to (\ref{tr1}), (\ref{tr2}) remain valid {\it cum grano salis}, see also \cite{BJMS10}.

\subsection{Characterization of $D_n$ \label{Unique}}
Note that the functional equations can be formulated for every spectral parameter $\la_i$ using the symmetry 
\begin{equation}
\check{R}_{i, i+1}(\la_{i}, \la_{i+1})D_n(\dots, \la_i,\la_{i+1},\dots)\check{R}_{i, i+1}(\la_{i+1}, \la_i)=D_n(\dots, \la_{i+1},\la_{i},\dots).\nonumber
\end{equation}
However, to give a (recursive) characterization of $D_n(\la_1, \dots, \la_n)$ it suffices to regard the functional equations for the last spectral parameter alone:

\noindent The $n$-site density operator $D_n(\la_1, \dots, \la_n)$ is uniquely characterized by the properties:
\begin{equation}
A_n(\la_1, \dots, \la_{n-1}, \nu_i)[D_n(\la_1, \dots, \la_{n-1}, \nu_i)]=D_n(\la_1, \dots, \la_{n-1}, \nu_i-1) \quad i=1\dots, N \label{dis_func}
\end{equation} 
and
\begin{equation}
\lim_{\lambda_n\rightarrow \infty} D_n(\lambda_1, \dots,\lambda_n)= \frac{1}{2} D_{n-1}(\lambda_1, \dots,\lambda_{n-1})\otimes I\label{Asymptotikbed}
\end{equation}
for generic values of the inhomogeneities $\nu_i$. The proof of this theorem is based on the functional dependence of the operator $D_n(\la_1, \dots, \la_n)$ on the $n$-th spectral parameter. 
For the XXX case each matrix element (\ref{Mat_el}) is a $n$-variate polynomial in the variables $\la_i$ of degree at most $N$ divided by a (known) polynomial of
the same type. For the XXZ case the same statement holds with respect to polynonials in the variables $q^{2 \la_i}$. As a consequence it suffices to show that the $N+1$ conditions given by equations (\ref{dis_func}) and (\ref{Asymptotikbed}) are independent. We state the proof in appendix \ref{Uniqueness}.
\section{The $n$-site density operator for the XXX chain}
For the XXX case we discuss the explicit solution for the inhomogeneous $n$-site density operator for finite Trotter number in full detail. The solution takes the same form as in the zero temperature case where only the function $\omega$ has to be replaced by its finite Trotter number counterpart. More precisely, we will show that the solution for arbitrary $n$ can be stated in terms of a single nearest neighbor correlator $\omega$ in the form 
\begin{equation}
D_n(\lambda_1,\lambda_2, \dots, \lambda_n)=\sum_{m=0}^{[n/2]}\sum_{I,J}\left(\prod_{p=1}^{m}\omega(\lambda_{I_p}, \lambda_{J_p})\right)f_{n,I,J}(\lambda_1,\lambda_2,\dots ,\lambda_n).\label{Sol_Dn}
\end{equation}
The structure coefficients
$f_{n,I,J}\in \mbox{End}\left(V^{\otimes n}\right)$ are the same as
constructed in \cite{BJMST04a} for the zero temperature case, quite explicit
formulas can be found in \cite{BST05,SST05}. These coefficients are sometimes called the \lq algebraic part\rq\, of the density matrix. The nearest neighbor correlator $\omega$ is called the \lq physical part\rq\, which aquires a temperature dependence. This simple correspondence of zero and finite temperature was conjectured in \cite{BGKS06} where the explicit factorization of multiple integral formulas for short distance correlators was carried out. Here we complete our proof of (\ref{Sol_Dn}) on the basis of discrete functional equations which is alternative to the algebraic proof by Jimbo, Miwa, and Smirnov \cite{JMS08}.
\subsection{The 2-site function $\omega$}\label{sec_om}
In the single site case the density matrix is completely fixed by symmetry, in
the two site case we find a linear combination of the identity operator and
the projector $Pr_s$ onto the singlet state (\ref{singlet})
\begin{equation}
 D_1(\lambda_1)=\frac{1}{2}\,\mbox{id},\quad
 D_2(\lambda_1,\lambda_2)=\frac{1}{4}\,\mbox{id}\otimes
 \mbox{id}+\frac{1}{3}\, \omega(\lambda_1,\lambda_2)
 \left(\frac{1}{2}\,\mbox{id}\otimes \mbox{id}- Pr_s\right). \label{1und2}
\end{equation}
Equation (\ref{1und2}) yields
\begin{equation}
\mbox{tr}\left(D_2(\mu, \la)\, Pr_s\right)=\frac{1}{2}-\omega(\mu, \la).\label{omega}
\end{equation}
From the properties of $D_2(\la_1,\la_2)$ we obtain important properties of
$\omega$ which are needed for the proof of equation (\ref{Sol_Dn}).
\begin{itemize}
\item [1)] The function $\om$ obeys the discrete functional equation\footnote{We use the abbreviation $\lambda_{ij}:=\lambda_i-\lambda_j$.} 
 \begin{equation}
\omega(\lambda_1,\lambda_2-1)=\frac{\lambda_{21}(\lambda_{21}-2)}{1-\lambda_{21}^2}\,\omega(\lambda_1,\lambda_2)-\frac{3}{2\left(1-\lambda_{21}^2\right)}\quad \mbox{for}\quad\lambda_2= \nu_1,\dots, \nu_N.\label{Funktional omega}
\end{equation}
\item [2)] The asymptotic behaviour is given by
\begin{equation}
 \omega(\lambda_1,\lambda_2)\xrightarrow[\lambda_2\rightarrow \infty]{} 0.\label{Asymptotik omega}
\end{equation}
 \item [3)] $\omega$ is symmetric under exchange of arguments: 
 \begin{equation}
 \omega(\la_1, \la_2)=\omega(\la_2, \la_1).\label{Symmetry omega}
 \end{equation} 
\item [4)] Furthermore $\Lambda_0(\la_1) \Lambda_0(\la_2) \,\omega(\la_1,
  \la_2)$ is a $2$ variate polynomial of degree $N$ with respect to $\la_1, \la_2$.
\end{itemize} 
An explicit expression of the function $\omega$ in terms of integral equations has been
obtained in \cite{BGKS06}. Here we present a derivation having
the advantage that all occurring functions can be obtained in terms of a
single auxiliary function. To this end, we introduce the modified quantum
transfer matrix in a larger Hilbert space
\begin{equation}
\tilde{V}_Q:=V_{\nu_1}\otimes V_{\nu_2}\otimes\dots \otimes V_{\nu_N}\otimes V_{\mu+1+\varepsilon}\otimes V_{\mu}.\nonumber
\end{equation}
Explicitly
\begin{equation*}
\tilde{T}(\la):=\mbox{tr}_{V_\la}\left(\check{R}_{N+2,N+3}(\la, \mu)\,\check{R}_{N+1,N+2}(\la, \mu+1+\varepsilon)\,\check{R}_{N,N+1}(\la, \nu_N)\, \cdots \,\check{R}_{1,2}(\la, \nu_1)\right).
\end{equation*}
Compared to $T(\la)$ the Trotter number of $\tilde{T}(\la)$ is enhanced by
two.  The parameter $\varepsilon$ is considered as a small deformation. For
$\varepsilon=0$ the leading eigenvector $\tilde{\Phi}_0$ of $\tilde{T}(\la)$
corresponds to the tensor product of the leading eigenvector ${\Phi}_0$ of
$T(\la)$ in the space $V_{\nu_1}\otimes\dots \otimes V_{\nu_N}$ with the 2-site $sl_2$-singlet (\ref{singlet}) in the two additional quantum spaces.  The
logarithmic derivative of the leading eigenvalue $\tilde{\Lambda}_0(\la)$ of
$\tilde{T}(\la)$ with respect to the small deformation parameter $\varepsilon$
evaluates to
\begin{multline}
\left(\frac{d}{d\varepsilon }\ln(\tilde{\Lambda}_0(\lambda))\right)|_{\varepsilon=0}
=\frac{\left<\tilde{\Phi}_0\right|\left(\frac{d}{d\varepsilon }\tilde{T}(\la)\right)\left|\tilde{\Phi}_0\right>}{\left<\tilde{\Phi}_0\right|\tilde{T}(\la)\left|\tilde{\Phi}_0\right>}|_{\varepsilon=0}\\
=\lim_{k\rightarrow \infty}\frac{\mbox{tr}_{\tilde{V}_Q}\left(\tilde{T}(\mu)\left[\tilde{T}(0)\right]^k\left(\frac{d}{d\varepsilon }\tilde{T}(\la)\right)\left[\tilde{T}(0)\right]^k\right) }{\mbox{tr}_{\tilde{V}_Q}\left(\tilde{T}(\mu)\left[\tilde{T}(0)\right]^k\tilde{T}(\la)\left[\tilde{T}(0)\right]^k \right) }|_{\varepsilon=0}\\
=\frac{\mbox{tr}_{V_\mu \otimes V_\lambda}\Bigl(D_2(\mu, \la)\, \check{R}(\la, \mu)\, \left(\frac{d}{d\varepsilon }\check{R}(\mu+\varepsilon, \la)\right)|_{\varepsilon=0}\Bigr)}{\mbox{tr}_{V_\mu \otimes V_\lambda}\left(D_2(\mu, \la)\, \check{R}(\la, \mu)\,\check{R}(\mu, \la)\right)}.\label{der_epsilon}
\end{multline}
Equation (\ref{der_epsilon}) connects the logarithmic derivative of the
leading eigenvalue $\tilde{\Lambda}_0(\la)$ to the expectation value of the
operator $\check{R}(\la, \mu)\, \left(\frac{d}{d\varepsilon
}\check{R}(\mu+\varepsilon, \la)\right)|_{\varepsilon=0}$ with respect to the
(unmodified) density operator $D_2(\mu, \la)$. A graphical illustration of equation (\ref{der_epsilon}) is given in Fig.~\ref{fig:omega}. 
\begin{figure}[h]
\centering
\resizebox{!}{5cm}{\input{FE_mod}}
\caption{Graphical illustration of the last equation in (\ref{der_epsilon}), see also Fig.~\ref{fig:R_dot}.}
\label{fig:omega}
\end{figure}
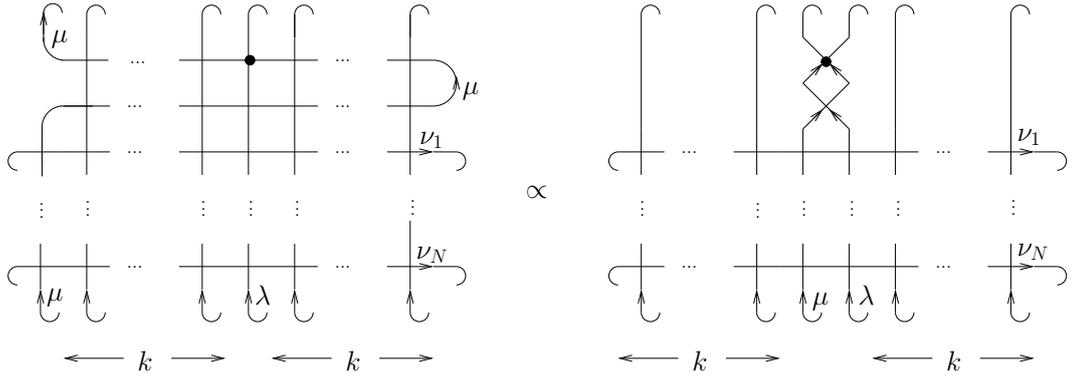
\begin{figure}[h]
\centering
\resizebox{!}{2.5cm}{\input{R_dot}}
\caption{The $R$-matrix symbol with central dot denotes the derivative $\left(\frac{d}{d\varepsilon }\check{R}(\mu+\varepsilon, \la)\right)|_{\varepsilon=0}$.}
\label{fig:R_dot}
\end{figure}
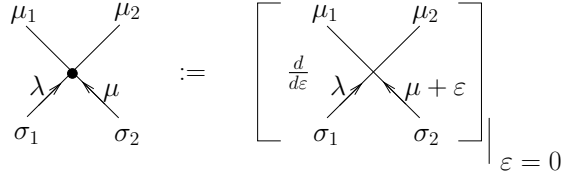

From definition (\ref{omega})
and $\mbox{tr}\left(D_2(\mu, \la)\right)=1$ we find
\begin{equation}
\left(\frac{d}{d\varepsilon }\ln(\tilde{\Lambda}_0(\lambda))\right)|_{\varepsilon=0}=\frac{1}{1-(\la-\mu)^2}\left((1+\la-\mu)-\frac{1}{2}+\omega(\mu, \la)\right).\label{om_Lam}
\end{equation}
For the leading eigenvalue $\tilde{\Lambda}_0(\lambda)$ the integral expression
\begin{multline}
\ln({\tilde{\Lambda}_0}({\la}))=\ln({\la}-{\mu}+1)+\ln({\la}-{\mu}-{\varepsilon}-1)\\
+\sum_{k=1}^{[N/2]}\ln\left({\la}-{\nu}_k+1\right)+\sum_{k=[N/2]+1}^{N}\ln({\la}-{\nu}_{k})\\-\frac{1}{2\pi i}\int_{{L}}\frac{1}{(\la-\om)(\la-\om+1)}\ln(1+\tilde{a}({\om}))d{\om}\label{L_1}
\end{multline}  
can be derived from the Bethe ansatz equations \cite{K92, K93}. The auxiliary function $\tilde{a}$ is determined by the integral equation
\begin{multline}
\ln(\tilde{a}(\la))=\ln\left(\frac{\la-\mu-1}{\la-\mu}\right)+\ln\left(\frac{\la-\mu-\varepsilon}{\la-\mu-\varepsilon-1}\right)\\
+\sum_{k=1}^{[N/2]}\ln\left(\frac{\la-\nu_k-1}{\la-\nu_k}\right)+\sum_{k=[N/2]+1}^N\ln\left(\frac{\la-\nu_{k}+1}{\la-\nu_{k}}\right)\\
+\frac{1}{2\pi i}\int_L\frac{2}{((\la-\om)^2-1)}\ln\left(1+\tilde{a}(\om)\right)d\om. \label{Int_a}
\end{multline}
The contour $L$ encloses the Bethe roots on the imaginary axis and also the
parameters $\mu$ and $\nu_i$ with $i\le [N/2]$, which we assume to be real and
close to zero. [Of course, by multiplying the spectral parameters by the
imaginary unit i the expressions can be written in terms of integrals along
the real axis.]  Setting $\varepsilon=0$ in equations (\ref{L_1}) and
(\ref{Int_a}) yields the description for the leading eigenvalue of the
unmodified quantum transfer matrix,
\begin{multline}
\ln({\Lambda}_0({\la}))=\ln(({\la}-{\mu})^2-1)+\sum_{k=1}^{[N/2]}\ln\left({\la}-{\nu}_k+1\right)+\sum_{k=[N/2]+1}^{N}\ln({\la}-{\nu}_{k})\\-\frac{1}{2\pi i}\int_{{L}}\frac{1}{(\la-\om)(\la-\om+1)}\ln(1+{a}({\om}))d{\om}, \nonumber
\end{multline} 
with auxiliary function $a(\la)=\tilde{a}(\la)_{|_{\varepsilon=0}}$ determined by
\begin{multline}
\ln(a(\la))=\sum_{k=1}^{[N/2]}\ln\left(\frac{\la-\nu_k-1}{\la-\nu_k}\right)+\sum_{k=[N/2]+1}^N\ln\left(\frac{\la-\nu_{k}+1}{\la-\nu_{k}}\right)\\
+\frac{1}{2\pi i}\int_L\frac{2}{((\la-\om)^2-1)}\ln\left(1+a(\om)\right)d\om.\label{a_Int}
\end{multline}
From equations (\ref{om_Lam}) and (\ref{L_1}) we obtain
\begin{equation}
\omega(\mu, \la)=\frac{1}{2}-\frac{\left(1-(\la-\mu)^2\right)}{2 \pi i}\int_L\frac{1}{(\la-\om)(\la-\om-1)}\frac{{G}({\omega}, {\mu})}{1+{a}({\omega)}}d{\omega}.\label{om_integral}
\end{equation}
The function $G$ introduced in \cite{GKS04a} is here obtained in terms of the modified auxiliary function
\begin{equation}
{G}({\la}, {\mu}):=\frac{d}{d{\varepsilon}}\ln(\tilde{a}({\la}))|_{{\varepsilon}=0}. \nonumber
\end{equation}
The linear integral equation 
\begin{equation}
 {G}({\la}, {\mu})=\frac{-1}{({\la}-{\mu})({\la}-{\mu}+1)}-\frac{1}{2\pi i}\int_{{L}}\frac{2}{((\la-\omega)^2-1)} \frac{{G}({\omega}, {\mu})}{1+{a}({\omega})}d{\omega}\label{G_Int}
 \end{equation}
follows from equation (\ref{Int_a}). Independent of our construction it is possible to take equations (\ref{a_Int})-(\ref{G_Int}) as definitions and read off the desired polynomial properties as well as (\ref{Funktional omega})-(\ref{Symmetry omega}).
\subsection{The algebraic part}\label{Solution_XXX}
We adopt the notations of \cite{BJMST04a, BJMST04b} to describe shortly the construction of the
algebraic part. It is convenient to consider the vector
\begin{equation}
h_n(\lambda_1, \dots, \lambda_n):=\left(D_n(\lambda_1, \dots, \lambda_n)\otimes I^{\otimes n}\right)({\bf s}_n),\label{dual}
\end{equation}
with the following numbering of the spaces:
\begin{equation}
h_n(\lambda_1, \dots, \lambda_n)\in V_1\otimes V_2\otimes\cdots\otimes V_n\otimes V_{\bar{n}}\otimes \cdots \otimes V_{\bar{2}} \otimes V_{\bar{1}}\nonumber
\end{equation}
The vector ${\bf s}_n\in V^{\otimes 2n}$ is defined by nested 2-site $sl_2$ singlets
\begin{equation}
{\bf s}_n:=\prod_{i=1}^n s_{i,\bar{i}},\quad \mbox{with}\quad s= v^+\otimes v^- -  v^-\otimes v^+ \label{Schachtelhalm}.
\end{equation}
Note that under the duality (\ref{dual}) the vector ${\bf s}_n$ corresponds to the identity operator on $n$ sites. Using the duality transformation and the crossing symmetry, $A_n \in \mbox{End}\left(V^{\otimes 2n}\right)$ takes the form
\begin{equation}
A_n^{(1)}=\left(\prod_{l=2}^{n} C_U^{-1}(\la_1, \la_l)\right) (-1)^n R_{\bar{1},\bar{2}}(\la_{1,2}-1)\cdots R_{\bar{1},\bar{n}}(\la_{1,n}-1)P_{1,\bar{1}} R_{1,n}(\la_{1,n})\cdots R_{1, 2}(\la_{1,2}). \label{A_2}
\end{equation} 
In this section we consider the functional equation with respect to $\la_1$. In \cite{BJMST04a} a family of operators
\begin{equation}
_{n}X_{n-2}^{(i,j)}(\la_1, \dots, \la_n)\in \mbox{Hom}\left(V^{\otimes 2(n-2)}, V^{\otimes 2n}\right),\quad 1\le i< j \le n,\nonumber
\end{equation}
called X-operators in the sequel, is constructed via transfer matrices whose
dimension of the auxiliary space has been continuously extended to $\mathbb{C}[\la]$. By construction the action of
$A_n$ on the X-operators is given by  
\begin{multline}
A_n^{(1)}(\lambda_1,\dots,\lambda_n)\,  {_{n}X^{(i,j)}_{n-2}}(\lambda_1, \dots, \lambda_n)=\\ \begin{cases} -{_{n}X^{(i,j)}_{n-2}}(\lambda_1-1, \la_2,\dots,\lambda_n)& \mbox{for}\; i=1,\\ {_{n}X^{(i,j)}_{n-2}}(\lambda_1-1, \la_2,\dots,\lambda_n)\, A_{n-2}^{(1)}(\lambda_1,\dots,\hat{\lambda_i},\dots,\hat{\lambda_j},\dots,\lambda_n)& \mbox{else.}\end{cases}\label{Funk_X}
\end{multline}
The symbol $\hat{\la}_{i}$ denotes the missing of the spectral parameter $\la_i$ in the sequence of arguments. The image of ${\bf s}_n$ evaluates to
\begin{multline}
A^{(1)}_{n}(\lambda_1,\dots,\lambda_n)({\bf s}_n)\\={\bf s}_n-6\, A^{(1)}_{n}(\lambda_1,\dots,\lambda_n)\left(\sum_{j=2}^n\left((1-\la_{j1}^2)\la_{j1}(\la_{j1}-2)\right)^{-1}{_nX^{(1,j)}_{n-2}}(\lambda_1,\dots,\lambda_n)({\bf s}_{n-2})\right).\label{Image_I}
\end{multline}
Furthermore, for pairwise distinct indices $i, j, k, l$ the commutation relation
\begin{multline}
{_{n}X^{(i,j)}_{n-2}}(\lambda_1,\dots,\lambda_n)\, {_{n-2}X^{(k',l')}_{n-4}}(\lambda_1, \dots,\hat{\lambda_i},\dots,\hat{\lambda_j},\dots,\lambda_n)\\
={_{n}X^{(k,l)}_{n-2}}(\lambda_1,\dots,\lambda_n)\, {_{n-2}X^{(i',j')}_{n-4}}(\lambda_1, \dots,\hat{\lambda_k},\dots,\hat{\lambda_l},\dots,\lambda_n),\label{Komm_X}
\end{multline}
holds, where $k'$ denotes the position of $\la_k$ in the sequence $\la_1, \dots, \hat{\la}_i,\dots,\hat{\la}_j,\dots, \la_n$.

\subsection{Solution for the $n$-site density operator}
The X-operators and the 2-site correlation function $\omega$, discussed in section \ref{sec_om}, are joined in the definition of $\Omega$-operators:
\begin{equation}
_{n}\Omega_{n-2}^{(i, j)}:=\frac{1}{\la_{ij}^2-1}\omega(\la_{i}, \la_{j})\,{_{n}X^{(i,j)}_{n-2}}(\lambda_{1},\dots,\lambda_{n}). \nonumber
\end{equation}
The result for $h_n$ is given by
\begin{equation}
h_n(\lambda_1,\dots,\lambda_n)=\sum_{m=0}^{[n/2]}\frac{(-1)^m}{2^{n-2m}}\sum_{I, J} \Omega_{K_1, (i_1, j_1)}\, \Omega_{K_2, (i_2, j_2)}\, \dots \, \Omega_{K_m, (i_m, j_m)}({\bf s}_{n-2m}).\label{Solution}
\end{equation}
For $K=\left\{k_1,\dots, k_m\right\}$ with $1\le k_1<\dots<k_m\le n$ the shorthand notation
\begin{equation}
\Omega_{K, (k_i, k_j)}:=_{m}\Omega_{m-2}^{(i, j)}(\lambda_{k_1},\dots,\lambda_{k_m}) \nonumber
\end{equation}
has been used. The inner sum on the \rhs of (\ref{Solution}) is performed over pairs of $m$-tuples, $I=(i_1,\dots, i_m),\;J=(j_1,\dots, j_m)$ of distinct elements of $K_1=\left\{1, 2, \dots, n \right\}$ under the condition  $i_1<\dots<i_m$ and $i_1<j_1,\dots, i_m<j_m$. The subsets are defined as $K_{p+1}:=K_p\setminus \left\{i_p, j_p\right\}$ for $p \geq 1$.
This expression for the inhomogeneous $n$-site density operator takes the same form as for the case of zero temperature. The only change is the replacement of the nearest neighbor correlator by the analogue function for the case of finite Trotter number. (This function $\omega$ determines the analytic properties of the expression.)
  
In order to prove that (\ref{Solution}) provides indeed the $n$-site density operator it is sufficient to check whether this expression is subject to the characterisation derived in section \ref{Unique}. From equations (\ref{Funk_X})-(\ref{Komm_X}) for the X-operators and the discrete functional equations (\ref{Funktional omega}) for $\omega$ it is straightforward to verify that the discrete functional equations (\ref{dis_func}) for the $n$-site density operator are fulfilled. To prove the asymptotic condition (\ref{Asymptotikbed}) one further needs \cite{BJMST04a} 
\begin{align}
&\lim_{\lambda_k\rightarrow \infty}\frac{1}{(\la_{ij}^2-1)}\,{_{n}X^{(i,j)}_{n-2}}(\lambda_{1},\dots,\lambda_{n})(s_{n-2}) = \mbox{const.} \quad\quad\forall \, k,\nonumber\\
&{_{n}X_{n-2}^{(i,j)}}(\lambda_1,\dots,\lambda_n)(s_{n-2})\xrightarrow[\lambda_1\rightarrow \infty]{} s_{1, \bar{1}}\;{_{n-2}X_{n-4}^{(i,j)}}(\lambda_2,\dots,\lambda_n)(s_{n-4}), i\neq 1\nonumber
\end{align}
The essential part of the proof that the \rhs of (\ref{Solution}) multiplied
by $\prod_{i=1}^n \Lambda_0(\la_i)$ is indeed an $n$-variate polynomial of
degree $N$ is to show that the additional poles coming along with the
$X$-operators vanish. This is achieved along the same lines as
in \cite{BJMST04b} by rewriting equation (\ref{Solution}) in exponential form
and using the symmetry property of $\omega$ (\ref{Symmetry omega}) to show
pairwise cancellation of residua.

\section{Discussion}
In this paper we showed how to derive the correlation functions of six-vertex
models on a lattice with infinite width and arbitrary height $N$. These systems
comprise as a special case, in the limit of infinite Trotter number
$N\to\infty$, the spin-1/2 Heisenberg chain in the thermodynamic limit at
finite temperature. Also, but not explicitly treated here, the general six-vertex model comprises the zero temperature Heisenberg chain on a finite ring \cite{DGHK07,SABGKTT11}.

The analysis was based on discrete functional equations (of
Knizhnik-Zamolodchikov type). For a certain set of $N$ many values of the
spectral parameter we showed the validity of the functional equations which
for zero temperature and field are known to hold for all complex
values. Despite the restricted validity, the discrete functional equations
suffice to uniquely identify the solution. For finite Trotter number $N$, the
dependence on the spectral parameter is polynomial of degree $N$. A formal
proof of the uniqueness was given.

A similar treatment of the correlation functions is possible for the
anisotropic spin-1/2 Heisenberg chain with finite temperature as well as
magnetic field, and even with a magnetic seam. Applications to related
Temperley-Lieb models and integrable higher spin-$S$ Heisenberg chains \cite{Ki01, GSS10} are
currently under investigation.

\vspace*{0.3 cm}
\noindent{\bf Acknowledgments:} 

The authors acknowledge valuable discussions with H. Boos, F. G\"ohmann, and J. Suzuki. B.A. was financially supported by the DFG research training group 1052 and by VolkswagenStiftung.

\begin{appendix}
\section{Solution for the $n$-site density operator of the XXZ chain}\label{XXZ_App}
For the XXZ case two independent nearest neighbor correlators $\omega$ and $\tilde{\omega}$ are necessary to state the result for $D_n(\la_1,\dots,\la_n)$. These two functions can be defined as follows
\begin{equation}
\omega(\la_1, \la_2):=\frac{1}{[\la_{12}-1]_q[\la_{12}+1]_q}\Bigl(\frac{1}{2}\ch(\eta)-\left<s(q)\right|D_2(\la_1, \la_2)\left|s(q)\right>\Bigr),\nonumber
\end{equation}
\begin{multline}
\tilde{\omega}(\la_1, \la_2):=\frac{1}{2 \ch(\eta)}\Bigl(\left<s(q)\right|D_2(\la_1, \la_2)\left|s(q)^t\right>-\left<s(q)^t\right|D_2(\la_1, \la_2)\left|s(q)\right>\Bigr)\\-\omega(\la_1, \la_2)\frac{\ch(\eta \la_{12})\sh(\eta \la_{12})}{\ch(\eta)\sh(\eta)}.\nonumber
\end{multline}
Here we used the notation 
\begin{equation*}
s(q):=q^{-1/2} v^+\otimes v^--q^{1/2}v^-\otimes v^+,\quad s(q)^t:=q^{-1/2} v^-\otimes v^+-q^{1/2}v^+\otimes v^-.\nonumber
\end{equation*}
The functions are the finite Trotter number analogues of the functions used in \cite{BJMST04b}. They are subject to the discrete functional equations
\begin{equation}
\begin{split}
\omega(\la_1, \la_2-1)&+\omega(\la_1, \la_2)+p(\la_1, \la_2)=0\\
\tilde{\omega}(\la_1, \la_2-1)&+\tilde{\omega}(\la_1, \la_2)+\omega(\la_1, \la_2)+\tilde{p}(\la_1, \la_2)+p(\la_1, \la_2)=0
\end{split}\nonumber
\end{equation} 
at $\la_2=\nu_1, \dots, \nu_N$.
\begin{equation}
\begin{split}
\tilde{p}(\la_1, \la_2)=&\frac{1}{2}\frac{1}{[\la_{12}-1]_q[\la_{12}-2]_q}-\frac{1}{4}\frac{[2]_q}{[\la_{12}-1]_q[\la_{12}+1]_q},\\
{p}(\la_1, \la_2)=&\frac{3}{4}\frac{1}{[\la_{12}]_q[\la_{12}-1]_q}-\frac{1}{4}\frac{[3]_q}{[\la_{12}-2]_q[\la_{12}+1]_q}.
\end{split}\nonumber
\end{equation}
The functions $\omega$ and $\tilde{\omega}$ are symmetric respectively antisymmetric with respect to exchange of arguments:
\begin{equation}
{\omega}(\la_1, \la_2)={\omega}(\la_2, \la_1),\quad\tilde{\omega}(\la_1, \la_2)=-\tilde{\omega}(\la_2, \la_1).\nonumber
\end{equation}
The vanishing of asymptotics follows from
\begin{equation}
\left<s(q)\right|D_2(\la_1, \la_2)\left|s(q)\right>\xrightarrow[\lambda_2\rightarrow \infty]{} \frac{1}{2}\ch(\eta)\nonumber
\end{equation}
respectively
\begin{equation}
\left<s(q)\right|D_2(\la_1, \la_2)\left|s(q)^t\right>\xrightarrow[\lambda_2\rightarrow \infty]{}-1,\quad\mbox{and}\quad \left<s(q)^t\right|D_2(\la_1, \la_2)\left|s(q)\right>\xrightarrow[\lambda_2\rightarrow \infty]{}-1.\nonumber
\end{equation}
The X-operators constructed in \cite{BJMST04b} take the form
\begin{equation}
_{n}X^{(i,j)}_{n-2}(\lambda_1, \dots, \lambda_n)=-\la_{i j}\,  _{n}\tilde{G}^{(i,j)}_{n-2}(\zeta_1, \dots, \zeta_n)+_{n}G^{(i,j)}_{n-2}(\zeta_1, \dots, \zeta_n),\nonumber
\end{equation}
with rational functions $_{n}\tilde{G}^{(i,j)}_{n-2}$ and $_{n}G^{(i,j)}_{n-2}$ with respect to the variables $\zeta_i:=q^{\la_i}$. The inhomogeneous $n$-site density operator for the XXZ chain is given by (\ref{Solution}) with the following definition of the $\Omega$-operators:
 \begin{equation}
_n\Omega_{n-2}^{(i, j)}:=\tilde{\omega}(\la_{i}, \la_{j})\,_{n}\tilde{G}^{(i,j)}_{n-2}(\zeta_{1},\dots,\zeta_{n})+\omega(\la_{i}, \la_{j})\,_{n}G^{(i,j)}_{n-2}(\zeta_{1},\dots,\zeta_{n}).\nonumber
\end{equation}

\section{Uniqueness of polynomial solutions to (\ref{dis_func}) and (\ref{Asymptotikbed})}{\label{Uniqueness}}
This appendix provides the proof that equations (\ref{dis_func}) and (\ref{Asymptotikbed}) uniquely determine the operator $D_n(\la_1,\dots,\la_n)$ for generic values of the inhomogeneities $\nu_1,\dots, \nu_N$ under the condition that the solution for $n-1$ sites is known. Recall from section \ref{Construction Dn} that the operator 
\begin{equation}
\Lambda_0(\la_n) D_n(\la_1,\dots,\la_n)\label{Lambda Dn}
\end{equation}
is a polynomial of degree $N$ with respect to the argument $\la_n$ for the XXX chain. The coefficients of this polynomial are linear operators on $V^{\otimes n}$, whose matrix entries are rational functions of $\la_1,\dots,\la_{n-1}$. For the XXZ chain the dependence of (\ref{Lambda Dn}) on $\la_n$ is of the form $\mathcal{F}_N(\la_n)$ with
\begin{equation}
\mathcal{F}_N(x):=\left\{q^{-Nx} \,P(q^{2x})\quad|\quad P\; \mbox{polynomial of degree}\; N\right\}.
\end{equation}
We present the proof of uniqueness for the XXZ case. For pairwise distinct $\nu_i$ a basis of the space $\mathcal{F}_N(x)$ is given by $(p_0, p_1,\dots, p_N)$ defined by
\begin{equation}
 p_0(x):=\prod_{i=1}^N [x-\nu_i]_q\;, \quad p_j(x):=q^{-(x-\nu_j)}\prod_{k=1, k\neq j}^N\frac{\left[x-\nu_k\right]_q}{\left[\nu_j-\nu_k\right]_q}\quad\mbox{for}\quad j=1,\dots, N. \nonumber
\end{equation}
Let $D_i$ be the coefficients of the operator (\ref{Lambda Dn}) with respect to the basis $(p_0,\dots, p_N)$ and $\tilde{D}_i$ the coefficients with respect to a second basis $(\tilde{p}_0,\dots,\tilde{p}_N)$ defined by
\begin{equation}
\tilde{p}_j(x):=p_j(x+1)\quad\mbox{for}\quad j=0, \dots, N. \nonumber
\end{equation} 
 By construction we have
\begin{equation}
D_i=\Lambda_0(\nu_i) D(\nu_i)\quad\mbox{and}\quad\tilde{D}_i=\Lambda_0(\nu_i-1)D(\nu_i-1)\nonumber
\end{equation} 
for $i=1, \dots, N$. The idea is to regard the coefficients $D_i$ as unknown quantities, to obtain a system of linear equations for them, and finally to show that this system has a unique solution in the generic case. The functional equations (\ref{dis_func}) are equivalent to the equations
\begin{equation}
A_n(\nu_i)\left [D_i\right
]=\frac{\Lambda_0(\nu_i)}{\Lambda_0(\nu_i-1)}\tilde{D}_i\quad\mbox{for} \quad i=1, \dots, N.\label{Lin1}
\end{equation}
The asymptotic condition (\ref{Asymptotikbed}) fixes the coefficients $D_0$ and $\tilde{D}_0$:
\begin{equation}
D_0=\tilde{D}_0=D_{n-1}(\lambda_1,\dots,\lambda_{n-1})\otimes\mbox{id}.
\end{equation}
 The subspace $\mathcal{F}_{N-1}$ is spanned by $\left(p_1,\dots, p_N\right)$ respectively $\left(\tilde{p}_1,\dots, \tilde{p}_N\right)$.
The elements of the matrix $W_N$ defined by 
\begin{equation}
{p}_i=\sum_{j=0}^{N}(W_N)_{ji}\, \tilde{p}_j,\nonumber
\end{equation}
are 
\begin{equation}
\left(W_N\right)_{0,l}=q^{-N}\, \delta_{0, l}\quad\mbox{for}\quad l=0,\dots, N,\quad \left(W_N\right)_{k,0}=\prod_{j=1}^N[\nu_k-\nu_j-1]_q\quad\mbox{for}\quad k=1,\dots, N,\nonumber
\end{equation}
and
\begin{equation}
(W_N)_{kl}=q^{\left(1+\nu_l-\nu_k\right)}\prod_{j=1, j\neq l}^N\frac{\left[\nu_k-\nu_j-1\right]_q}{\left[\nu_l-\nu_j\right]_q}\quad \mbox{for}\quad  k,l=1, \dots N. \nonumber
\end{equation}
Equations (\ref{Lin1}) yield a system of linear equations for the coefficients $\tilde{D}_i$: 
\begin{equation}
\frac{\Lambda_0(\nu_i-1)}{\Lambda_0(\nu_i)} A_n(\nu_i)\left [\sum_{j=0}^N (W_N)_{ij}\tilde{D}_j\right
]=\tilde{D}_i   \quad\mbox{for} \quad i=1, \dots, N.\label{LinSys}
\end{equation}
We choose an arbitrary basis in the vector space $\mbox{End}(V^{\otimes n})$ and denote by $\tilde{\mathbb{D}}_i$ the coordinate vector of $\tilde{{D}}_i$ and by  $\mathbb{A}(\nu_i)$ the describing matrix of the linear map ${A}_n(\nu_i)$. Equations (\ref{LinSys}) yield a system of linear equations for the vectors $\tilde{\mathbb{D}}_i,$ $i=1,\dots,N$, where the constant term is determined by the  (known) vector $\tilde{\mathbb{D}}_0$.
The corresponding homogeneous linear system reads
\begin{equation}
\vec{\tilde{\mathbb{D}}}=
\mathbb{L}\, \mathbb{A}\,\mathbb{W}
\,\vec{\tilde{\mathbb{D}}}.\label{homogenes LGS}
\end{equation}
The $N \times N$ block matrices $\mathbb{L}\,, \mathbb{A},$ and $\mathbb{W}$ with blocks of size $n^2 \times n^2$ have the entries
\begin{equation}
\left(\mathbb{W}\right )_{i,j}:=\left( W_N\right )_{i,j}\mathds{1},\quad \left(\mathbb{L}\right)_{i,j}:=\frac{\Lambda_0(\nu_i-1)}{\Lambda_0(\nu_i)}\,\delta_{i,j}\,\mathds{1},\nonumber
\end{equation}
and
\begin{equation}
\left(\mathbb{A}\right)_{i,j}:=\delta_{i,j}\,\mathbb{A}(\nu_i)\quad\mbox{for}\quad i,j=1,\dots,N. \nonumber
\end{equation}
The solution of (\ref{homogenes LGS}) is unique iff
\begin{equation}
\mbox{det}\left(\mathbb{L}\,\mathbb{A}\,\mathbb{W}-\mathds{1}\right)\neq 0.\nonumber
\end{equation}
We make the ansatz
\begin{equation}
 \nu_{2l-1}=n_l+d+\frac{1}{2},\quad \nu_{2l}=n_l-d+\frac{1}{2}\quad\mbox{for}\quad l=1, \dots, [N/2]\label{ansatz nu}
\end{equation}
with a constant parameter $d$. The Bethe roots $n_l$ are determined by the Bethe ansatz equations
\begin{equation}
\prod_{j=1}^{N}\frac{\left[n_l-\nu_j+1\right]_q}{\left[n_l-\nu_j\right]_q}=(-1)\prod_{k=1}^{[N/2]}\frac{\left[n_l-n_k+1\right]_q}{\left[n_l-n_k-1\right]_q}\quad\mbox{for}\;l=1,\dots, [N/2]\nonumber
\end{equation}
for the leading eigenvalue 
\begin{equation}
\Lambda_0(\lambda)=\,\prod_{j=1}^{N}\left[1+\lambda-\nu_j\right]_q\prod_{l=1}^{[N/2]}\frac{\left[n_l-\lambda+1\right]_q}{\left[n_l-\lambda\right]_q}+\prod_{j=1}^{N}\left[\lambda-\nu_j\right]\prod_{l=1}^{[N/2]}\frac{\left[\lambda-n_l+1\right]_q}{\left[\lambda-n_l\right]_q}.\nonumber
\end{equation}
We regard the limit
\begin{equation}
 n_l\rightarrow \infty\quad\mbox{under the condition}\quad \left |n_l-n_k\right |_{l\neq k}\rightarrow \infty\quad \forall\; l,k.\nonumber
\end{equation}
In this limit the blocks on the diagonal of $\mathbb{A}$ are constant and have the only eigenvalues $\pm 1$. Furthermore $\mathbb{W}\, \mathbb{L}$ commutes with $\mathbb{A}$. It thus suffices to show that all eigenvalues of $\mathbb{W}\, \mathbb{L}$ are different from $\pm 1$. We find
\begin{equation}
 \frac{\Lambda_0(\nu_{2l-1}-1)}{\Lambda_0(\nu_{2l-1})}= \frac{\left[1+2d\right]_q}{\left[1-2d\right]_q}\quad,\quad \frac{\Lambda_0(\nu_{2l}-1)}{\Lambda_0(\nu_{2l})}= \frac{\left[1-2d\right]_q}{\left[1+2d\right]_q}\nonumber
\end{equation}
and 
\begin{align}
 (W_N)_{2l-1,2l-1}&=\frac{\left[2d-1\right]_q}{\left[ 2d\right]_q}q& (W_N)_{2l-1,2l}=&\frac{1}{\left[ 2d\right]_q} q^{-( 2d-1)}\nonumber\\(W_N)_{2l,2l-1}&=\frac{-1}{\left[ 2d\right]_q}q^{( 2d+1)}&(W_N)_{2l,2l}=&\frac{\left[2d+1\right]_q}{\left[ 2d\right]_q}q \nonumber
\end{align}
for $l=1,\dots, [N/2]$. The remaining elements of $W_N$ vanish. The parameter $d$ can easily be chosen in such a way that the eigenvalues of the $2 \times 2$ matrices on the diagonal of $\mathbb{W}\, \mathbb{L}$ are not $\pm 1$.

\section{Functional equations in the presence of a magnetic field and disorder}\label{Funk_alpha}
The functional equations (\ref{AD}) for the inhomogeneous $n$-site density
operator of the Heisenberg chain (\ref{XXX-Hamiltonian}) hold also in the
presence of a magnetic field $h$, which for the 6-vertex model corresponds to
a horizontal seam $\left(e^{(h \beta/2) \sigma_z}\right)^{\otimes
  L}$. For the XXZ case ($\Delta\neq 1$) the introduction of a
disorder parameter $\alpha$ brings forth a fermionic structure on the space of local operators \cite{BJMST06b, BJMST08a} and consequently a structural simplification of the solution. For our approach, the additional interaction terms do not present fundamental obstacles. Discrete functional equations can be derived and analytical properties established. The solution to these equations is again unique.
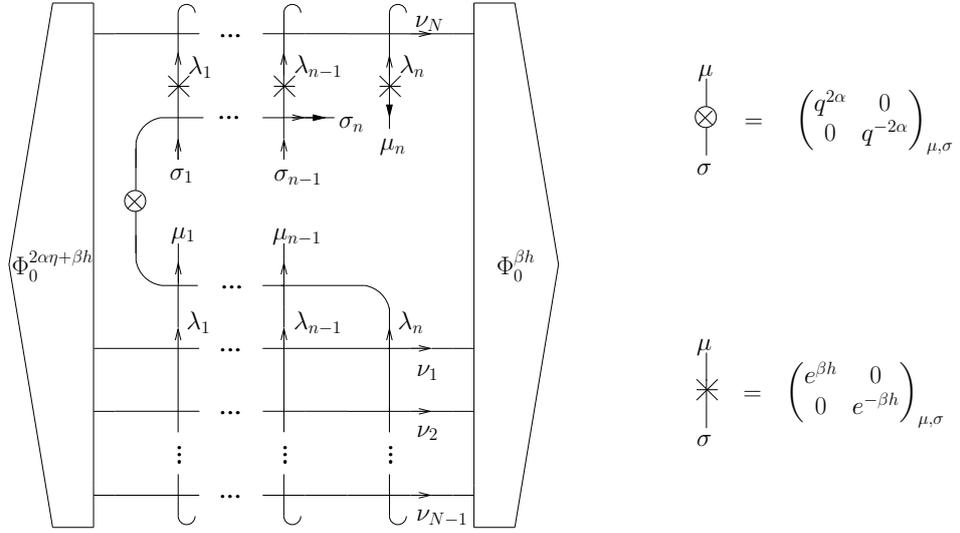
\begin{figure}[h!]
\centering
\resizebox{!}{7 cm}{\input{A_alpha}}
\caption{Graphical illustration of the matrix element
  $\left(A_n\left[D_n(\la_1, \dots, \la_n)\right]\right)^{\mu_1 \dots
    \mu_n}_{\sigma_1 \dots \sigma_n}$. For the normalization of the density matrix
  we would have to multiply it with $\prod_{i=1}^n \Lambda^{(\beta h)}_0(\la_i)$ and the inverse of the constant (\ref{C_A_alpha}).}
\label{fig:A_op_al}
\end{figure}

The generalized inhomogeneous $n$-site density operator in the presence of a magnetic field and disorder parameter is defined as
\begin{equation}
\begin{split}
D_n(\la_1, \dots, \la_n):=&\lim_{k\rightarrow\infty}\frac{\mbox{tr}_{V_Q}\left(\left[T^{(2\alpha \eta+h \beta)}(0)\right]^k \mathcal{T}^{(h \beta)}(\la_1)\mathcal{T}^{(h \beta)}(\la_2) \cdots  \mathcal{T}^{(h \beta)}(\la_n) \left[T^{(h \beta)}(0)\right]^{k}\right)}{\mbox{tr}_{V_Q}\left(\left[T^{(2\alpha \eta+h \beta)}(0)\right]^k{T}^{(h \beta)}(\la_1){T}^{(h \beta)}(\la_2)\cdots {T}^{(h \beta)}(\la_n) \left[T^{(h \beta)}(0)\right]^k\right)}\\
=&\frac{\left<\Phi_0^{(2\alpha \eta+h \beta)}\right| \mathcal{T}^{(h \beta)}(\la_1)\mathcal{T}^{(h \beta)}(\la_2) \cdots  \mathcal{T}^{(h \beta)}(\la_n) \left|\Phi_0^{(h \beta)}\right>}{{\Lambda}_0^{(h \beta)}(\la_1){\Lambda_0}^{(h \beta)}(\la_2)\cdots {\Lambda_0}^{(h \beta)}(\la_n) },
\end{split}\nonumber
\end{equation}
where $\Phi_0^{(\kappa)}$ denotes the eigenvector corresponding to the leading eigenvalue $\Lambda_0^{\kappa}$ of the twisted transfer matrix
\begin{equation*}
T^{(\kappa)}(\la):=\mbox{tr}_{V_\la}\left(\check{R}_{N,N+1}(\la, \nu_N)\, \cdots \,\check{R}_{1,2}(\la, \nu_1)e^{(\kappa/2) \sigma_z}_1\right),\quad \kappa \in \mathbb{C}.\nonumber
\end{equation*}
The discrete functional equations hold for the generalized $n$-site density operator, if the definition of the operator $A_n$ includes the disorder field in the way shown in figure (\ref{fig:A_op_al}) and the constant $C_A$ is set to
\begin{equation}
C_A:=\frac{\Lambda_0^{(h \beta)}(\la_n)}{\Lambda_0^{(2\alpha \eta+h \beta)}(\la_n)}\prod_{l=1}^{n-1}C_U^{-1}(\lambda_n, \lambda_l).\label{C_A_alpha}
\end{equation}
Here we do not present the solution to the functional equations but like to refer the reader to \cite{JMS08, BoGo09}.

\end{appendix}

\bibliographystyle{unsrt}

\end{document}

%% file: R_ind.tex
\begin{picture}(0,0)%
\includegraphics{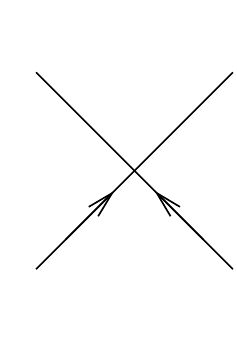}%
\end{picture}%
\setlength{\unitlength}{4144sp}%
\begingroup\makeatletter\ifx\SetFigFont\undefined%
\gdef\SetFigFont#1#2#3#4#5{%
  \reset@font\fontsize{#1}{#2pt}%
  \fontfamily{#3}\fontseries{#4}\fontshape{#5}%
  \selectfont}%
\fi\endgroup%
\begin{picture}(1077,1546)(4111,-4211)
\put(4141,-4111){\makebox(0,0)[lb]{\smash{{\SetFigFont{17}{20.4}{\familydefault}{\mddefault}{\updefault}{\color[rgb]{0,0,0}$\sigma_1$}%
}}}}
\put(5131,-4111){\makebox(0,0)[lb]{\smash{{\SetFigFont{17}{20.4}{\familydefault}{\mddefault}{\updefault}{\color[rgb]{0,0,0}$\sigma_2$}%
}}}}
\put(4126,-2896){\makebox(0,0)[lb]{\smash{{\SetFigFont{17}{20.4}{\familydefault}{\mddefault}{\updefault}{\color[rgb]{0,0,0}$\mu_1$}%
}}}}
\put(5116,-2896){\makebox(0,0)[lb]{\smash{{\SetFigFont{17}{20.4}{\familydefault}{\mddefault}{\updefault}{\color[rgb]{0,0,0}$\mu_2$}%
}}}}
\put(4201,-3646){\makebox(0,0)[lb]{\smash{{\SetFigFont{17}{20.4}{\familydefault}{\mddefault}{\updefault}{\color[rgb]{0,0,0}$\lambda_2$}%
}}}}
\put(4996,-3661){\makebox(0,0)[lb]{\smash{{\SetFigFont{17}{20.4}{\familydefault}{\mddefault}{\updefault}{\color[rgb]{0,0,0}$\lambda_1$}%
}}}}
\end{picture}%

%% file: YBE_UNI_ind_fig.tex
\begin{picture}(0,0)%
\includegraphics{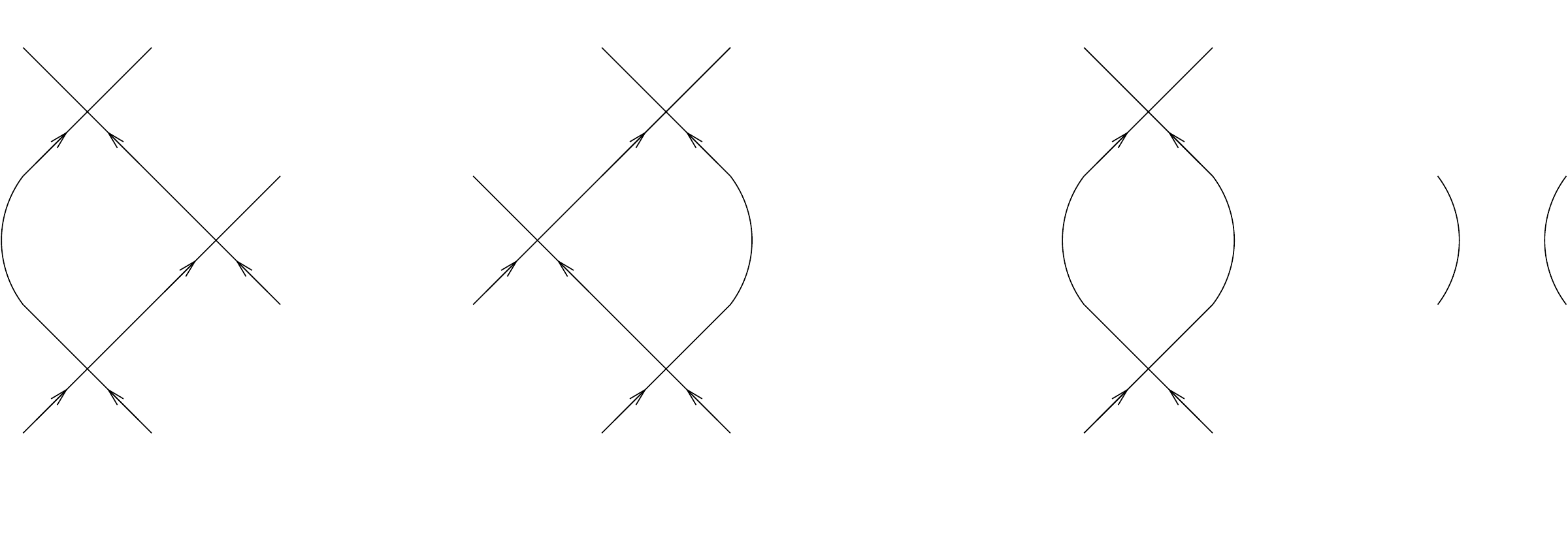}%
\end{picture}%
\setlength{\unitlength}{4144sp}%
\begingroup\makeatletter\ifx\SetFigFont\undefined%
\gdef\SetFigFont#1#2#3#4#5{%
  \reset@font\fontsize{#1}{#2pt}%
  \fontfamily{#3}\fontseries{#4}\fontshape{#5}%
  \selectfont}%
\fi\endgroup%
\begin{picture}(10969,3846)(965,-4486)
\put(10897,-1771){\makebox(0,0)[lb]{\smash{{\SetFigFont{17}{20.4}{\familydefault}{\mddefault}{\updefault}{\color[rgb]{0,0,0}$\mu_1$}%
}}}}
\put(11842,-1786){\makebox(0,0)[lb]{\smash{{\SetFigFont{17}{20.4}{\familydefault}{\mddefault}{\updefault}{\color[rgb]{0,0,0}$\mu_2$}%
}}}}
\put(10876,-2986){\makebox(0,0)[lb]{\smash{{\SetFigFont{17}{20.4}{\familydefault}{\mddefault}{\updefault}{\color[rgb]{0,0,0}$\sigma_1$}%
}}}}
\put(11896,-2986){\makebox(0,0)[lb]{\smash{{\SetFigFont{17}{20.4}{\familydefault}{\mddefault}{\updefault}{\color[rgb]{0,0,0}$\sigma_2$}%
}}}}
\put(991,-3886){\makebox(0,0)[lb]{\smash{{\SetFigFont{17}{20.4}{\familydefault}{\mddefault}{\updefault}{\color[rgb]{0,0,0}$\sigma_1$}%
}}}}
\put(1981,-3886){\makebox(0,0)[lb]{\smash{{\SetFigFont{17}{20.4}{\familydefault}{\mddefault}{\updefault}{\color[rgb]{0,0,0}$\sigma_2$}%
}}}}
\put(2842,-1786){\makebox(0,0)[lb]{\smash{{\SetFigFont{17}{20.4}{\familydefault}{\mddefault}{\updefault}{\color[rgb]{0,0,0}$\mu_3$}%
}}}}
\put(1957,-871){\makebox(0,0)[lb]{\smash{{\SetFigFont{17}{20.4}{\familydefault}{\mddefault}{\updefault}{\color[rgb]{0,0,0}$\mu_2$}%
}}}}
\put(997,-871){\makebox(0,0)[lb]{\smash{{\SetFigFont{17}{20.4}{\familydefault}{\mddefault}{\updefault}{\color[rgb]{0,0,0}$\mu_1$}%
}}}}
\put(1021,-3436){\makebox(0,0)[lb]{\smash{{\SetFigFont{17}{20.4}{\familydefault}{\mddefault}{\updefault}{\color[rgb]{0,0,0}$\lambda_3$}%
}}}}
\put(1876,-3451){\makebox(0,0)[lb]{\smash{{\SetFigFont{17}{20.4}{\familydefault}{\mddefault}{\updefault}{\color[rgb]{0,0,0}$\lambda_2$}%
}}}}
\put(1921,-2521){\makebox(0,0)[lb]{\smash{{\SetFigFont{17}{20.4}{\familydefault}{\mddefault}{\updefault}{\color[rgb]{0,0,0}$\lambda_3$}%
}}}}
\put(2776,-2521){\makebox(0,0)[lb]{\smash{{\SetFigFont{17}{20.4}{\familydefault}{\mddefault}{\updefault}{\color[rgb]{0,0,0}$\lambda_1$}%
}}}}
\put(1021,-1636){\makebox(0,0)[lb]{\smash{{\SetFigFont{17}{20.4}{\familydefault}{\mddefault}{\updefault}{\color[rgb]{0,0,0}$\lambda_2$}%
}}}}
\put(1876,-1636){\makebox(0,0)[lb]{\smash{{\SetFigFont{17}{20.4}{\familydefault}{\mddefault}{\updefault}{\color[rgb]{0,0,0}$\lambda_1$}%
}}}}
\put(2926,-2986){\makebox(0,0)[lb]{\smash{{\SetFigFont{17}{20.4}{\familydefault}{\mddefault}{\updefault}{\color[rgb]{0,0,0}$\sigma_3$}%
}}}}
\put(5026,-3886){\makebox(0,0)[lb]{\smash{{\SetFigFont{17}{20.4}{\familydefault}{\mddefault}{\updefault}{\color[rgb]{0,0,0}$\sigma_2$}%
}}}}
\put(6046,-3886){\makebox(0,0)[lb]{\smash{{\SetFigFont{17}{20.4}{\familydefault}{\mddefault}{\updefault}{\color[rgb]{0,0,0}$\sigma_3$}%
}}}}
\put(5056,-3421){\makebox(0,0)[lb]{\smash{{\SetFigFont{17}{20.4}{\familydefault}{\mddefault}{\updefault}{\color[rgb]{0,0,0}$\lambda_2$}%
}}}}
\put(5911,-3436){\makebox(0,0)[lb]{\smash{{\SetFigFont{17}{20.4}{\familydefault}{\mddefault}{\updefault}{\color[rgb]{0,0,0}$\lambda_1$}%
}}}}
\put(4141,-2986){\makebox(0,0)[lb]{\smash{{\SetFigFont{17}{20.4}{\familydefault}{\mddefault}{\updefault}{\color[rgb]{0,0,0}$\sigma_1$}%
}}}}
\put(4171,-2536){\makebox(0,0)[lb]{\smash{{\SetFigFont{17}{20.4}{\familydefault}{\mddefault}{\updefault}{\color[rgb]{0,0,0}$\lambda_3$}%
}}}}
\put(5011,-2521){\makebox(0,0)[lb]{\smash{{\SetFigFont{17}{20.4}{\familydefault}{\mddefault}{\updefault}{\color[rgb]{0,0,0}$\lambda_1$}%
}}}}
\put(4147,-1771){\makebox(0,0)[lb]{\smash{{\SetFigFont{17}{20.4}{\familydefault}{\mddefault}{\updefault}{\color[rgb]{0,0,0}$\mu_1$}%
}}}}
\put(5056,-1621){\makebox(0,0)[lb]{\smash{{\SetFigFont{17}{20.4}{\familydefault}{\mddefault}{\updefault}{\color[rgb]{0,0,0}$\lambda_3$}%
}}}}
\put(5911,-1621){\makebox(0,0)[lb]{\smash{{\SetFigFont{17}{20.4}{\familydefault}{\mddefault}{\updefault}{\color[rgb]{0,0,0}$\lambda_2$}%
}}}}
\put(5047,-871){\makebox(0,0)[lb]{\smash{{\SetFigFont{17}{20.4}{\familydefault}{\mddefault}{\updefault}{\color[rgb]{0,0,0}$\mu_2$}%
}}}}
\put(5992,-871){\makebox(0,0)[lb]{\smash{{\SetFigFont{17}{20.4}{\familydefault}{\mddefault}{\updefault}{\color[rgb]{0,0,0}$\mu_3$}%
}}}}
\put(8416,-3886){\makebox(0,0)[lb]{\smash{{\SetFigFont{17}{20.4}{\familydefault}{\mddefault}{\updefault}{\color[rgb]{0,0,0}$\sigma_1$}%
}}}}
\put(9406,-3886){\makebox(0,0)[lb]{\smash{{\SetFigFont{17}{20.4}{\familydefault}{\mddefault}{\updefault}{\color[rgb]{0,0,0}$\sigma_2$}%
}}}}
\put(8416,-3436){\makebox(0,0)[lb]{\smash{{\SetFigFont{17}{20.4}{\familydefault}{\mddefault}{\updefault}{\color[rgb]{0,0,0}$\lambda_2$}%
}}}}
\put(9301,-3436){\makebox(0,0)[lb]{\smash{{\SetFigFont{17}{20.4}{\familydefault}{\mddefault}{\updefault}{\color[rgb]{0,0,0}$\lambda_1$}%
}}}}
\put(8422,-871){\makebox(0,0)[lb]{\smash{{\SetFigFont{17}{20.4}{\familydefault}{\mddefault}{\updefault}{\color[rgb]{0,0,0}$\mu_1$}%
}}}}
\put(9382,-886){\makebox(0,0)[lb]{\smash{{\SetFigFont{17}{20.4}{\familydefault}{\mddefault}{\updefault}{\color[rgb]{0,0,0}$\mu_2$}%
}}}}
\put(8431,-1636){\makebox(0,0)[lb]{\smash{{\SetFigFont{17}{20.4}{\familydefault}{\mddefault}{\updefault}{\color[rgb]{0,0,0}$\lambda_1$}%
}}}}
\put(9301,-1621){\makebox(0,0)[lb]{\smash{{\SetFigFont{17}{20.4}{\familydefault}{\mddefault}{\updefault}{\color[rgb]{0,0,0}$\lambda_2$}%
}}}}
\put(9826,-2401){\makebox(0,0)[lb]{\smash{{\SetFigFont{17}{20.4}{\familydefault}{\mddefault}{\updefault}{\color[rgb]{0,0,0}$=$}%
}}}}
\put(10381,-2446){\makebox(0,0)[lb]{\smash{{\SetFigFont{17}{20.4}{\familydefault}{\mddefault}{\updefault}{\color[rgb]{0,0,0}$C_U$}%
}}}}
\put(3466,-2341){\makebox(0,0)[lb]{\smash{{\SetFigFont{17}{20.4}{\familydefault}{\mddefault}{\updefault}{\color[rgb]{0,0,0}$=$}%
}}}}
\end{picture}%

%% file: ANF_CROSS_ind.tex
\begin{picture}(0,0)%
\includegraphics{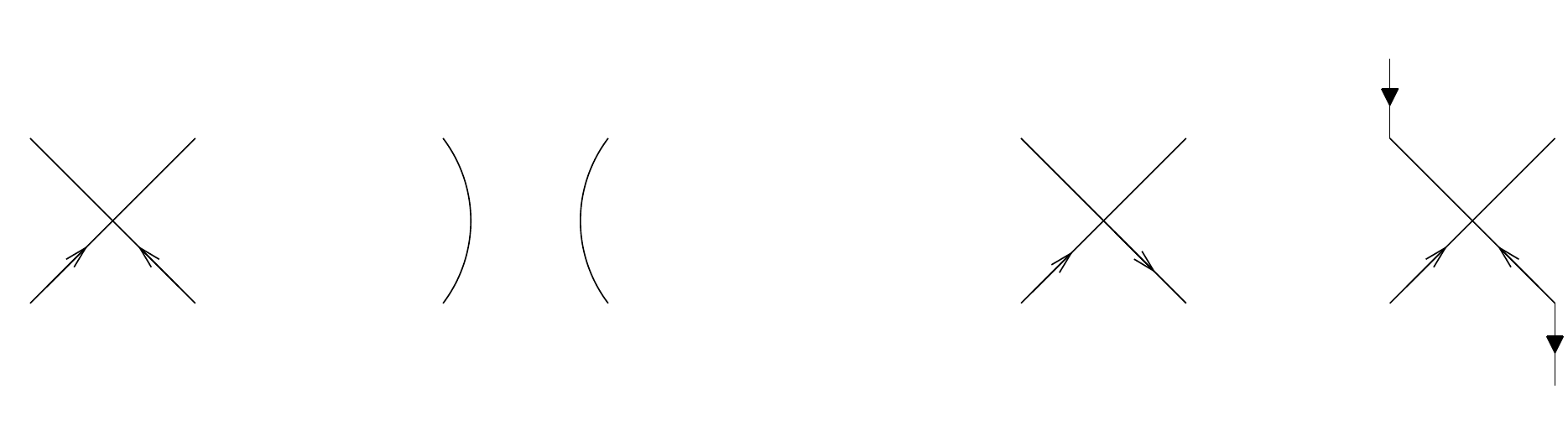}%
\end{picture}%
\setlength{\unitlength}{4144sp}%
\begingroup\makeatletter\ifx\SetFigFont\undefined%
\gdef\SetFigFont#1#2#3#4#5{%
  \reset@font\fontsize{#1}{#2pt}%
  \fontfamily{#3}\fontseries{#4}\fontshape{#5}%
  \selectfont}%
\fi\endgroup%
\begin{picture}(8532,2416)(961,-8261)
\put(3247,-6496){\makebox(0,0)[lb]{\smash{{\SetFigFont{17}{20.4}{\familydefault}{\mddefault}{\updefault}{\color[rgb]{0,0,0}$\mu_1$}%
}}}}
\put(4192,-6511){\makebox(0,0)[lb]{\smash{{\SetFigFont{17}{20.4}{\familydefault}{\mddefault}{\updefault}{\color[rgb]{0,0,0}$\mu_2$}%
}}}}
\put(3226,-7711){\makebox(0,0)[lb]{\smash{{\SetFigFont{17}{20.4}{\familydefault}{\mddefault}{\updefault}{\color[rgb]{0,0,0}$\sigma_1$}%
}}}}
\put(4246,-7711){\makebox(0,0)[lb]{\smash{{\SetFigFont{17}{20.4}{\familydefault}{\mddefault}{\updefault}{\color[rgb]{0,0,0}$\sigma_2$}%
}}}}
\put(997,-6496){\makebox(0,0)[lb]{\smash{{\SetFigFont{17}{20.4}{\familydefault}{\mddefault}{\updefault}{\color[rgb]{0,0,0}$\mu_1$}%
}}}}
\put(1957,-6511){\makebox(0,0)[lb]{\smash{{\SetFigFont{17}{20.4}{\familydefault}{\mddefault}{\updefault}{\color[rgb]{0,0,0}$\mu_2$}%
}}}}
\put(976,-7711){\makebox(0,0)[lb]{\smash{{\SetFigFont{17}{20.4}{\familydefault}{\mddefault}{\updefault}{\color[rgb]{0,0,0}$\sigma_1$}%
}}}}
\put(1981,-7711){\makebox(0,0)[lb]{\smash{{\SetFigFont{17}{20.4}{\familydefault}{\mddefault}{\updefault}{\color[rgb]{0,0,0}$\sigma_2$}%
}}}}
\put(1021,-7246){\makebox(0,0)[lb]{\smash{{\SetFigFont{17}{20.4}{\familydefault}{\mddefault}{\updefault}{\color[rgb]{0,0,0}$\lambda_1$}%
}}}}
\put(1876,-7246){\makebox(0,0)[lb]{\smash{{\SetFigFont{17}{20.4}{\familydefault}{\mddefault}{\updefault}{\color[rgb]{0,0,0}$\lambda_1$}%
}}}}
\put(6397,-6496){\makebox(0,0)[lb]{\smash{{\SetFigFont{17}{20.4}{\familydefault}{\mddefault}{\updefault}{\color[rgb]{0,0,0}$\mu_1$}%
}}}}
\put(7357,-6511){\makebox(0,0)[lb]{\smash{{\SetFigFont{17}{20.4}{\familydefault}{\mddefault}{\updefault}{\color[rgb]{0,0,0}$\mu_2$}%
}}}}
\put(6376,-7711){\makebox(0,0)[lb]{\smash{{\SetFigFont{17}{20.4}{\familydefault}{\mddefault}{\updefault}{\color[rgb]{0,0,0}$\sigma_1$}%
}}}}
\put(7381,-7711){\makebox(0,0)[lb]{\smash{{\SetFigFont{17}{20.4}{\familydefault}{\mddefault}{\updefault}{\color[rgb]{0,0,0}$\sigma_2$}%
}}}}
\put(7276,-7246){\makebox(0,0)[lb]{\smash{{\SetFigFont{17}{20.4}{\familydefault}{\mddefault}{\updefault}{\color[rgb]{0,0,0}$\lambda_1$}%
}}}}
\put(6406,-7246){\makebox(0,0)[lb]{\smash{{\SetFigFont{17}{20.4}{\familydefault}{\mddefault}{\updefault}{\color[rgb]{0,0,0}$\lambda_2$}%
}}}}
\put(2656,-7111){\makebox(0,0)[lb]{\smash{{\SetFigFont{17}{20.4}{\familydefault}{\mddefault}{\updefault}{\color[rgb]{0,0,0}$=$}%
}}}}
\put(7846,-7111){\makebox(0,0)[lb]{\smash{{\SetFigFont{17}{20.4}{\familydefault}{\mddefault}{\updefault}{\color[rgb]{0,0,0}$=$}%
}}}}
\put(8401,-7711){\makebox(0,0)[lb]{\smash{{\SetFigFont{17}{20.4}{\familydefault}{\mddefault}{\updefault}{\color[rgb]{0,0,0}$\sigma_1$}%
}}}}
\put(8416,-7261){\makebox(0,0)[lb]{\smash{{\SetFigFont{17}{20.4}{\familydefault}{\mddefault}{\updefault}{\color[rgb]{0,0,0}$\lambda_2$}%
}}}}
\put(9376,-8161){\makebox(0,0)[lb]{\smash{{\SetFigFont{17}{20.4}{\familydefault}{\mddefault}{\updefault}{\color[rgb]{0,0,0}$\sigma_2$}%
}}}}
\put(9301,-7246){\makebox(0,0)[lb]{\smash{{\SetFigFont{17}{20.4}{\familydefault}{\mddefault}{\updefault}{\color[rgb]{0,0,0}$\lambda_1-1$}%
}}}}
\put(8392,-6076){\makebox(0,0)[lb]{\smash{{\SetFigFont{17}{20.4}{\familydefault}{\mddefault}{\updefault}{\color[rgb]{0,0,0}$\mu_1$}%
}}}}
\put(9352,-6496){\makebox(0,0)[lb]{\smash{{\SetFigFont{17}{20.4}{\familydefault}{\mddefault}{\updefault}{\color[rgb]{0,0,0}$\mu_2$}%
}}}}
\end{picture}%

%% file: ID_SING_alternativ.tex
\begin{picture}(0,0)%
\includegraphics{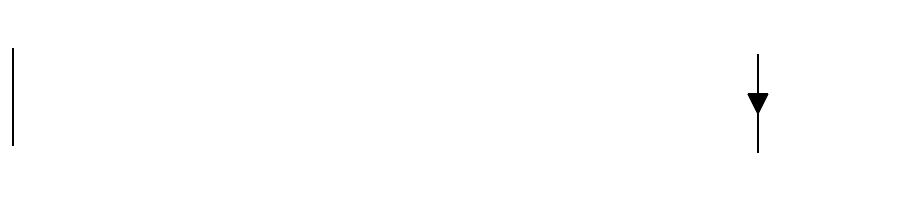}%
\end{picture}%
\setlength{\unitlength}{4144sp}%
\begingroup\makeatletter\ifx\SetFigFont\undefined%
\gdef\SetFigFont#1#2#3#4#5{%
  \reset@font\fontsize{#1}{#2pt}%
  \fontfamily{#3}\fontseries{#4}\fontshape{#5}%
  \selectfont}%
\fi\endgroup%
\begin{picture}(4185,940)(6466,-7544)
\put(9871,-6766){\makebox(0,0)[lb]{\smash{{\SetFigFont{12}{14.4}{\familydefault}{\mddefault}{\updefault}{\color[rgb]{0,0,0}$\mu$}%
}}}}
\put(9886,-7471){\makebox(0,0)[lb]{\smash{{\SetFigFont{12}{14.4}{\familydefault}{\mddefault}{\updefault}{\color[rgb]{0,0,0}$\sigma$}%
}}}}
\put(6736,-7051){\makebox(0,0)[lb]{\smash{{\SetFigFont{12}{14.4}{\familydefault}{\mddefault}{\updefault}{\color[rgb]{0,0,0}$=$}%
}}}}
\put(6481,-6751){\makebox(0,0)[lb]{\smash{{\SetFigFont{12}{14.4}{\familydefault}{\mddefault}{\updefault}{\color[rgb]{0,0,0}$\mu$}%
}}}}
\put(10201,-7066){\makebox(0,0)[lb]{\smash{{\SetFigFont{12}{14.4}{\familydefault}{\mddefault}{\updefault}{\color[rgb]{0,0,0}$=$}%
}}}}
\put(6481,-7426){\makebox(0,0)[lb]{\smash{{\SetFigFont{12}{14.4}{\familydefault}{\mddefault}{\updefault}{\color[rgb]{0,0,0}$\sigma$}%
}}}}
\put(7156,-7036){\makebox(0,0)[lb]{\smash{{\SetFigFont{12}{14.4}{\familydefault}{\mddefault}{\updefault}{\color[rgb]{0,0,0}$\begin{pmatrix}1&0\\0&1\end{pmatrix}_{\mu, \sigma}$}%
}}}}
\put(10636,-7066){\makebox(0,0)[lb]{\smash{{\SetFigFont{12}{14.4}{\familydefault}{\mddefault}{\updefault}{\color[rgb]{0,0,0}$\begin{pmatrix}0&-1\\1 &0\end{pmatrix}_{\mu, \sigma}$}%
}}}}
\end{picture}%

%% file: Zahler_ind.tex
\begin{picture}(0,0)%
\includegraphics{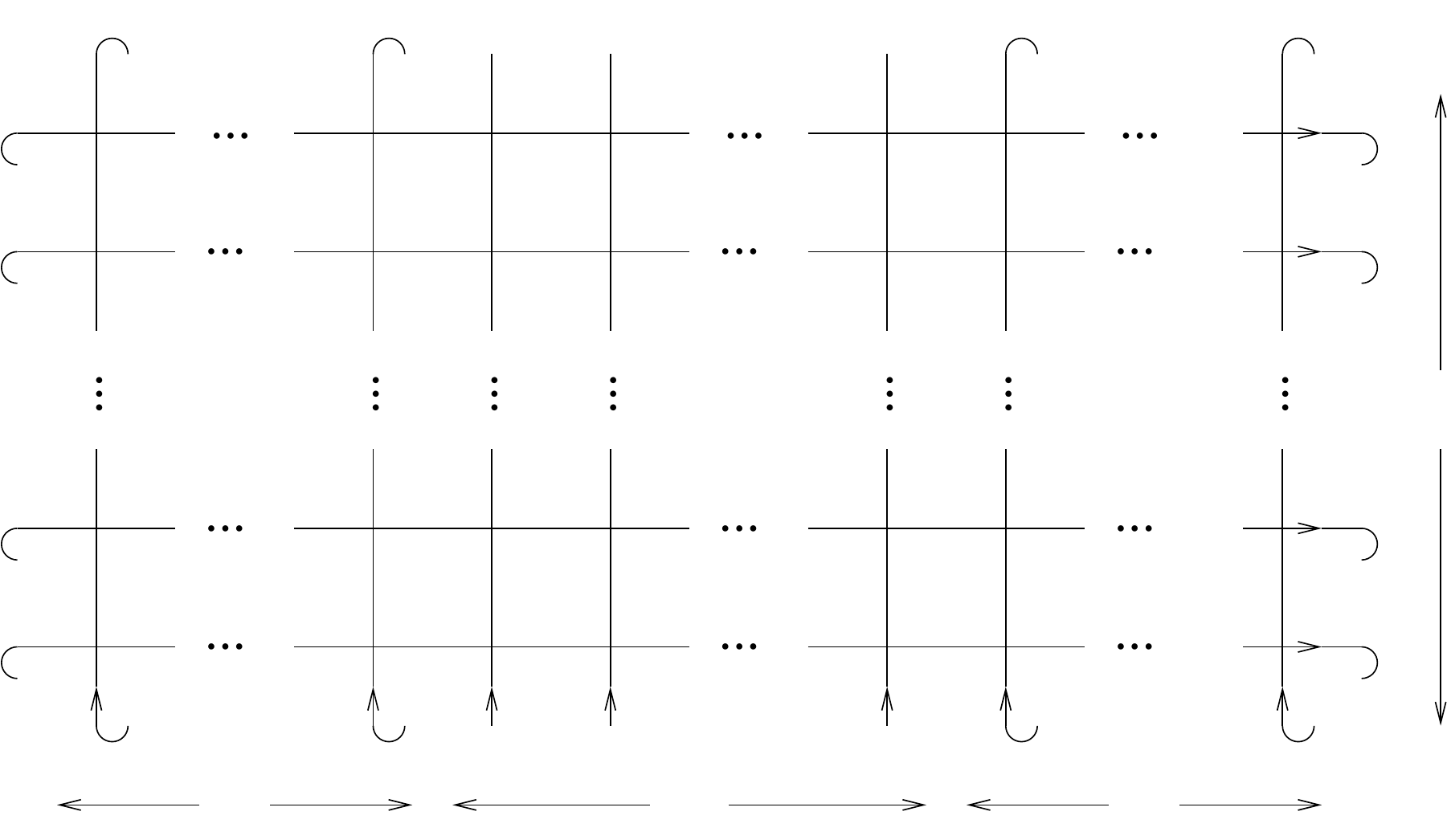}%
\end{picture}%
\setlength{\unitlength}{4144sp}%
\begingroup\makeatletter\ifx\SetFigFont\undefined%
\gdef\SetFigFont#1#2#3#4#5{%
  \reset@font\fontsize{#1}{#2pt}%
  \fontfamily{#3}\fontseries{#4}\fontshape{#5}%
  \selectfont}%
\fi\endgroup%
\begin{picture}(8241,4777)(2377,-9482)
\put(3541,-8401){\makebox(0,0)[lb]{\smash{{\SetFigFont{20}{24.0}{\rmdefault}{\mddefault}{\updefault}{\color[rgb]{0,0,0}...}%
}}}}
\put(3541,-7726){\makebox(0,0)[lb]{\smash{{\SetFigFont{20}{24.0}{\rmdefault}{\mddefault}{\updefault}{\color[rgb]{0,0,0}...}%
}}}}
\put(3541,-6151){\makebox(0,0)[lb]{\smash{{\SetFigFont{20}{24.0}{\rmdefault}{\mddefault}{\updefault}{\color[rgb]{0,0,0}...}%
}}}}
\put(3571,-5491){\makebox(0,0)[lb]{\smash{{\SetFigFont{20}{24.0}{\rmdefault}{\mddefault}{\updefault}{\color[rgb]{0,0,0}...}%
}}}}
\put(2956,-7066){\rotatebox{90.0}{\makebox(0,0)[lb]{\smash{{\SetFigFont{20}{24.0}{\rmdefault}{\mddefault}{\updefault}{\color[rgb]{0,0,0}...}%
}}}}}
\put(4531,-7066){\rotatebox{90.0}{\makebox(0,0)[lb]{\smash{{\SetFigFont{20}{24.0}{\rmdefault}{\mddefault}{\updefault}{\color[rgb]{0,0,0}...}%
}}}}}
\put(9706,-7066){\rotatebox{90.0}{\makebox(0,0)[lb]{\smash{{\SetFigFont{20}{24.0}{\rmdefault}{\mddefault}{\updefault}{\color[rgb]{0,0,0}...}%
}}}}}
\put(8746,-5491){\makebox(0,0)[lb]{\smash{{\SetFigFont{20}{24.0}{\rmdefault}{\mddefault}{\updefault}{\color[rgb]{0,0,0}...}%
}}}}
\put(5206,-7066){\rotatebox{90.0}{\makebox(0,0)[lb]{\smash{{\SetFigFont{20}{24.0}{\rmdefault}{\mddefault}{\updefault}{\color[rgb]{0,0,0}...}%
}}}}}
\put(6496,-5491){\makebox(0,0)[lb]{\smash{{\SetFigFont{20}{24.0}{\rmdefault}{\mddefault}{\updefault}{\color[rgb]{0,0,0}...}%
}}}}
\put(6466,-8401){\makebox(0,0)[lb]{\smash{{\SetFigFont{20}{24.0}{\rmdefault}{\mddefault}{\updefault}{\color[rgb]{0,0,0}...}%
}}}}
\put(6466,-7726){\makebox(0,0)[lb]{\smash{{\SetFigFont{20}{24.0}{\rmdefault}{\mddefault}{\updefault}{\color[rgb]{0,0,0}...}%
}}}}
\put(6466,-6151){\makebox(0,0)[lb]{\smash{{\SetFigFont{20}{24.0}{\rmdefault}{\mddefault}{\updefault}{\color[rgb]{0,0,0}...}%
}}}}
\put(5881,-7066){\rotatebox{90.0}{\makebox(0,0)[lb]{\smash{{\SetFigFont{20}{24.0}{\rmdefault}{\mddefault}{\updefault}{\color[rgb]{0,0,0}...}%
}}}}}
\put(7456,-7066){\rotatebox{90.0}{\makebox(0,0)[lb]{\smash{{\SetFigFont{20}{24.0}{\rmdefault}{\mddefault}{\updefault}{\color[rgb]{0,0,0}...}%
}}}}}
\put(8716,-8401){\makebox(0,0)[lb]{\smash{{\SetFigFont{20}{24.0}{\rmdefault}{\mddefault}{\updefault}{\color[rgb]{0,0,0}...}%
}}}}
\put(8716,-7726){\makebox(0,0)[lb]{\smash{{\SetFigFont{20}{24.0}{\rmdefault}{\mddefault}{\updefault}{\color[rgb]{0,0,0}...}%
}}}}
\put(8716,-6151){\makebox(0,0)[lb]{\smash{{\SetFigFont{20}{24.0}{\rmdefault}{\mddefault}{\updefault}{\color[rgb]{0,0,0}...}%
}}}}
\put(8131,-7066){\rotatebox{90.0}{\makebox(0,0)[lb]{\smash{{\SetFigFont{20}{24.0}{\rmdefault}{\mddefault}{\updefault}{\color[rgb]{0,0,0}...}%
}}}}}
\put(5071,-9061){\makebox(0,0)[lb]{\smash{{\SetFigFont{17}{20.4}{\familydefault}{\mddefault}{\updefault}{\color[rgb]{0,0,0}$\sigma_1$}%
}}}}
\put(5746,-9061){\makebox(0,0)[lb]{\smash{{\SetFigFont{17}{20.4}{\familydefault}{\mddefault}{\updefault}{\color[rgb]{0,0,0}$\sigma_2$}%
}}}}
\put(7321,-9061){\makebox(0,0)[lb]{\smash{{\SetFigFont{17}{20.4}{\familydefault}{\mddefault}{\updefault}{\color[rgb]{0,0,0}$\sigma_n$}%
}}}}
\put(5926,-8761){\makebox(0,0)[lb]{\smash{{\SetFigFont{17}{20.4}{\familydefault}{\mddefault}{\updefault}{\color[rgb]{0,0,0}$\lambda_2$}%
}}}}
\put(7501,-8776){\makebox(0,0)[lb]{\smash{{\SetFigFont{17}{20.4}{\familydefault}{\mddefault}{\updefault}{\color[rgb]{0,0,0}$\lambda_n$}%
}}}}
\put(5251,-8746){\makebox(0,0)[lb]{\smash{{\SetFigFont{17}{20.4}{\familydefault}{\mddefault}{\updefault}{\color[rgb]{0,0,0}$\lambda_1$}%
}}}}
\put(9766,-7621){\makebox(0,0)[lb]{\smash{{\SetFigFont{17}{20.4}{\familydefault}{\mddefault}{\updefault}{\color[rgb]{0,0,0}$\nu_2$}%
}}}}
\put(9751,-6046){\makebox(0,0)[lb]{\smash{{\SetFigFont{17}{20.4}{\familydefault}{\mddefault}{\updefault}{\color[rgb]{0,0,0}$\nu_{N-1}$}%
}}}}
\put(9751,-5371){\makebox(0,0)[lb]{\smash{{\SetFigFont{17}{20.4}{\familydefault}{\mddefault}{\updefault}{\color[rgb]{0,0,0}$\nu_N$}%
}}}}
\put(10411,-7156){\makebox(0,0)[lb]{\smash{{\SetFigFont{17}{20.4}{\familydefault}{\mddefault}{\updefault}{\color[rgb]{0,0,0}$N$}%
}}}}
\put(3646,-9391){\makebox(0,0)[lb]{\smash{{\SetFigFont{17}{20.4}{\familydefault}{\mddefault}{\updefault}{\color[rgb]{0,0,0}$k$}%
}}}}
\put(8821,-9376){\makebox(0,0)[lb]{\smash{{\SetFigFont{17}{20.4}{\familydefault}{\mddefault}{\updefault}{\color[rgb]{0,0,0}$k$}%
}}}}
\put(5071,-4936){\makebox(0,0)[lb]{\smash{{\SetFigFont{17}{20.4}{\familydefault}{\mddefault}{\updefault}{\color[rgb]{0,0,0}$\mu_1$}%
}}}}
\put(5746,-4936){\makebox(0,0)[lb]{\smash{{\SetFigFont{17}{20.4}{\familydefault}{\mddefault}{\updefault}{\color[rgb]{0,0,0}$\mu_2$}%
}}}}
\put(7321,-4936){\makebox(0,0)[lb]{\smash{{\SetFigFont{17}{20.4}{\familydefault}{\mddefault}{\updefault}{\color[rgb]{0,0,0}$\mu_n$}%
}}}}
\put(6226,-9376){\makebox(0,0)[lb]{\smash{{\SetFigFont{17}{20.4}{\familydefault}{\mddefault}{\updefault}{\color[rgb]{0,0,0}$n$}%
}}}}
\put(9751,-8296){\makebox(0,0)[lb]{\smash{{\SetFigFont{17}{20.4}{\familydefault}{\mddefault}{\updefault}{\color[rgb]{0,0,0}$\nu_1$}%
}}}}
\end{picture}%

%% file: Def_A.tex
\begin{picture}(0,0)%
\includegraphics{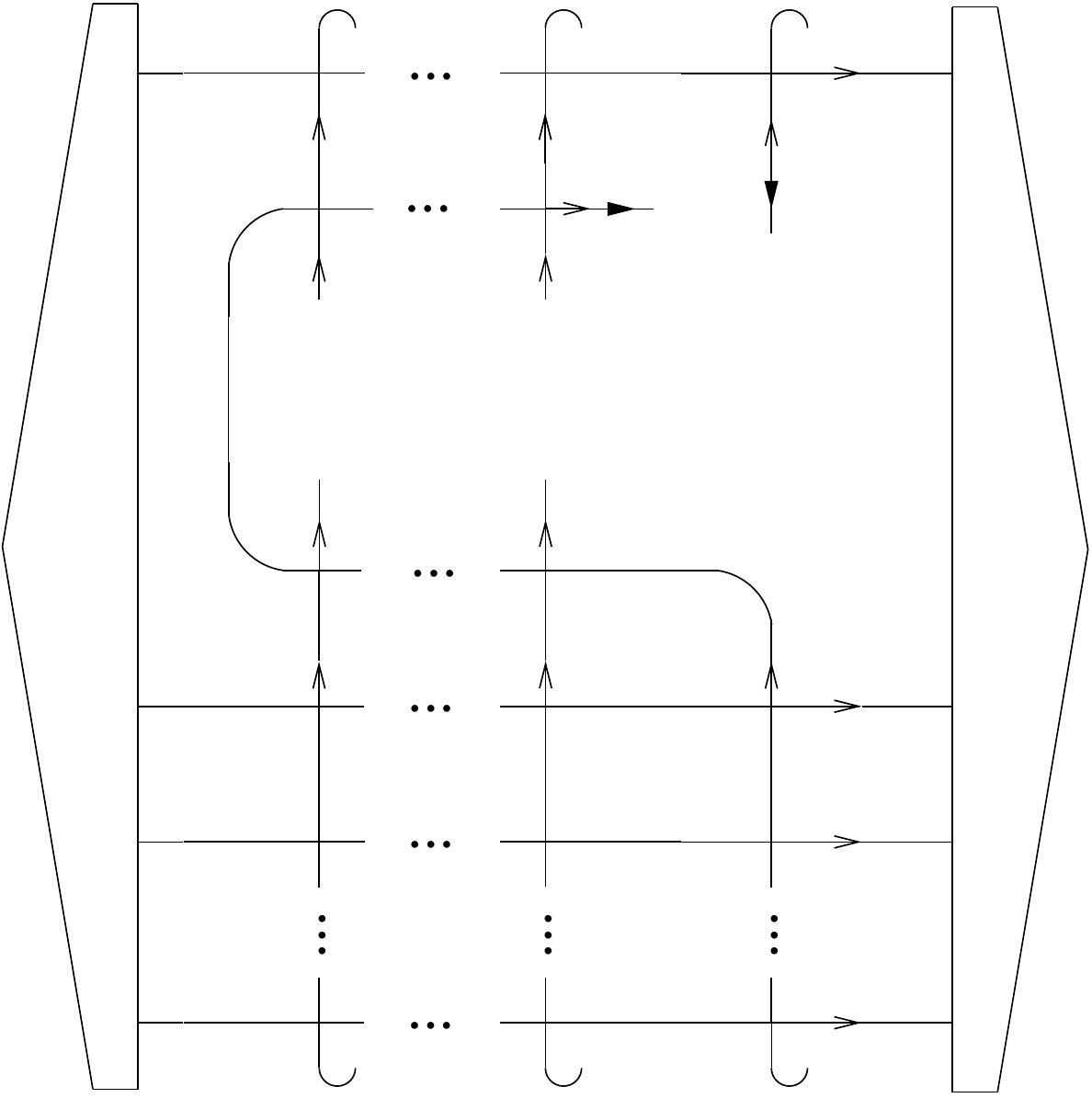}%
\end{picture}%
\setlength{\unitlength}{4144sp}%
\begingroup\makeatletter\ifx\SetFigFont\undefined%
\gdef\SetFigFont#1#2#3#4#5{%
  \reset@font\fontsize{#1}{#2pt}%
  \fontfamily{#3}\fontseries{#4}\fontshape{#5}%
  \selectfont}%
\fi\endgroup%
\begin{picture}(5424,5453)(439,-5382)
\put(676,-2746){\makebox(0,0)[lb]{\smash{{\SetFigFont{17}{20.4}{\familydefault}{\mddefault}{\updefault}{\color[rgb]{0,0,0}$\Phi_0$}%
}}}}
\put(5356,-2761){\makebox(0,0)[lb]{\smash{{\SetFigFont{17}{20.4}{\familydefault}{\mddefault}{\updefault}{\color[rgb]{0,0,0}$\Phi_0$}%
}}}}
\put(2116,-3226){\makebox(0,0)[lb]{\smash{{\SetFigFont{17}{20.4}{\familydefault}{\mddefault}{\updefault}{\color[rgb]{0,0,0}$\lambda_1$}%
}}}}
\put(3256,-3226){\makebox(0,0)[lb]{\smash{{\SetFigFont{17}{20.4}{\familydefault}{\mddefault}{\updefault}{\color[rgb]{0,0,0}$\lambda_{n-1}$}%
}}}}
\put(4364,-3227){\makebox(0,0)[lb]{\smash{{\SetFigFont{17}{20.4}{\familydefault}{\mddefault}{\updefault}{\color[rgb]{0,0,0}$\la_n$}%
}}}}
\put(2478,-2789){\makebox(0,0)[lb]{\smash{{\SetFigFont{20}{24.0}{\rmdefault}{\mddefault}{\updefault}{\color[rgb]{0,0,0}...}%
}}}}
\put(2463,-3464){\makebox(0,0)[lb]{\smash{{\SetFigFont{20}{24.0}{\rmdefault}{\mddefault}{\updefault}{\color[rgb]{0,0,0}...}%
}}}}
\put(2448,-975){\makebox(0,0)[lb]{\smash{{\SetFigFont{20}{24.0}{\rmdefault}{\mddefault}{\updefault}{\color[rgb]{0,0,0}...}%
}}}}
\put(2463,-315){\makebox(0,0)[lb]{\smash{{\SetFigFont{20}{24.0}{\rmdefault}{\mddefault}{\updefault}{\color[rgb]{0,0,0}...}%
}}}}
\put(1936,-1621){\makebox(0,0)[lb]{\smash{{\SetFigFont{17}{20.4}{\familydefault}{\mddefault}{\updefault}{\color[rgb]{0,0,0}$\sigma_1$}%
}}}}
\put(3046,-1636){\makebox(0,0)[lb]{\smash{{\SetFigFont{17}{20.4}{\familydefault}{\mddefault}{\updefault}{\color[rgb]{0,0,0}$\sigma_{n-1}$}%
}}}}
\put(1951,-2251){\makebox(0,0)[lb]{\smash{{\SetFigFont{17}{20.4}{\familydefault}{\mddefault}{\updefault}{\color[rgb]{0,0,0}$\mu_1$}%
}}}}
\put(3031,-2266){\makebox(0,0)[lb]{\smash{{\SetFigFont{17}{20.4}{\familydefault}{\mddefault}{\updefault}{\color[rgb]{0,0,0}$\mu_{n-1}$}%
}}}}
\put(2463,-4139){\makebox(0,0)[lb]{\smash{{\SetFigFont{20}{24.0}{\rmdefault}{\mddefault}{\updefault}{\color[rgb]{0,0,0}...}%
}}}}
\put(2462,-5040){\makebox(0,0)[lb]{\smash{{\SetFigFont{20}{24.0}{\rmdefault}{\mddefault}{\updefault}{\color[rgb]{0,0,0}...}%
}}}}
\put(2027,-4456){\rotatebox{270.0}{\makebox(0,0)[lb]{\smash{{\SetFigFont{20}{24.0}{\rmdefault}{\mddefault}{\updefault}{\color[rgb]{0,0,0}...}%
}}}}}
\put(3152,-4456){\rotatebox{270.0}{\makebox(0,0)[lb]{\smash{{\SetFigFont{20}{24.0}{\rmdefault}{\mddefault}{\updefault}{\color[rgb]{0,0,0}...}%
}}}}}
\put(4277,-4456){\rotatebox{270.0}{\makebox(0,0)[lb]{\smash{{\SetFigFont{20}{24.0}{\rmdefault}{\mddefault}{\updefault}{\color[rgb]{0,0,0}...}%
}}}}}
\put(2133,-720){\makebox(0,0)[lb]{\smash{{\SetFigFont{17}{20.4}{\familydefault}{\mddefault}{\updefault}{\color[rgb]{0,0,0}$\lambda_1$}%
}}}}
\put(3258,-705){\makebox(0,0)[lb]{\smash{{\SetFigFont{17}{20.4}{\familydefault}{\mddefault}{\updefault}{\color[rgb]{0,0,0}$\lambda_{n-1}$}%
}}}}
\put(4381,-706){\makebox(0,0)[lb]{\smash{{\SetFigFont{17}{20.4}{\familydefault}{\mddefault}{\updefault}{\color[rgb]{0,0,0}$\la_n$}%
}}}}
\put(4548,-211){\makebox(0,0)[lb]{\smash{{\SetFigFont{17}{20.4}{\familydefault}{\mddefault}{\updefault}{\color[rgb]{0,0,0}$\nu_N$}%
}}}}
\put(4563,-3735){\makebox(0,0)[lb]{\smash{{\SetFigFont{17}{20.4}{\familydefault}{\mddefault}{\updefault}{\color[rgb]{0,0,0}$\nu_1$}%
}}}}
\put(4562,-5282){\makebox(0,0)[lb]{\smash{{\SetFigFont{17}{20.4}{\familydefault}{\mddefault}{\updefault}{\color[rgb]{0,0,0}$\nu_{N-1}$}%
}}}}
\put(4563,-4380){\makebox(0,0)[lb]{\smash{{\SetFigFont{17}{20.4}{\familydefault}{\mddefault}{\updefault}{\color[rgb]{0,0,0}$\nu_2$}%
}}}}
\put(4171,-1306){\makebox(0,0)[lb]{\smash{{\SetFigFont{17}{20.4}{\familydefault}{\mddefault}{\updefault}{\color[rgb]{0,0,0}$\mu_n$}%
}}}}
\put(3736,-1066){\makebox(0,0)[lb]{\smash{{\SetFigFont{17}{20.4}{\familydefault}{\mddefault}{\updefault}{\color[rgb]{0,0,0}$\sigma_n$}%
}}}}
\end{picture}%

%% file: A_plus.tex
\begin{picture}(0,0)%
\includegraphics{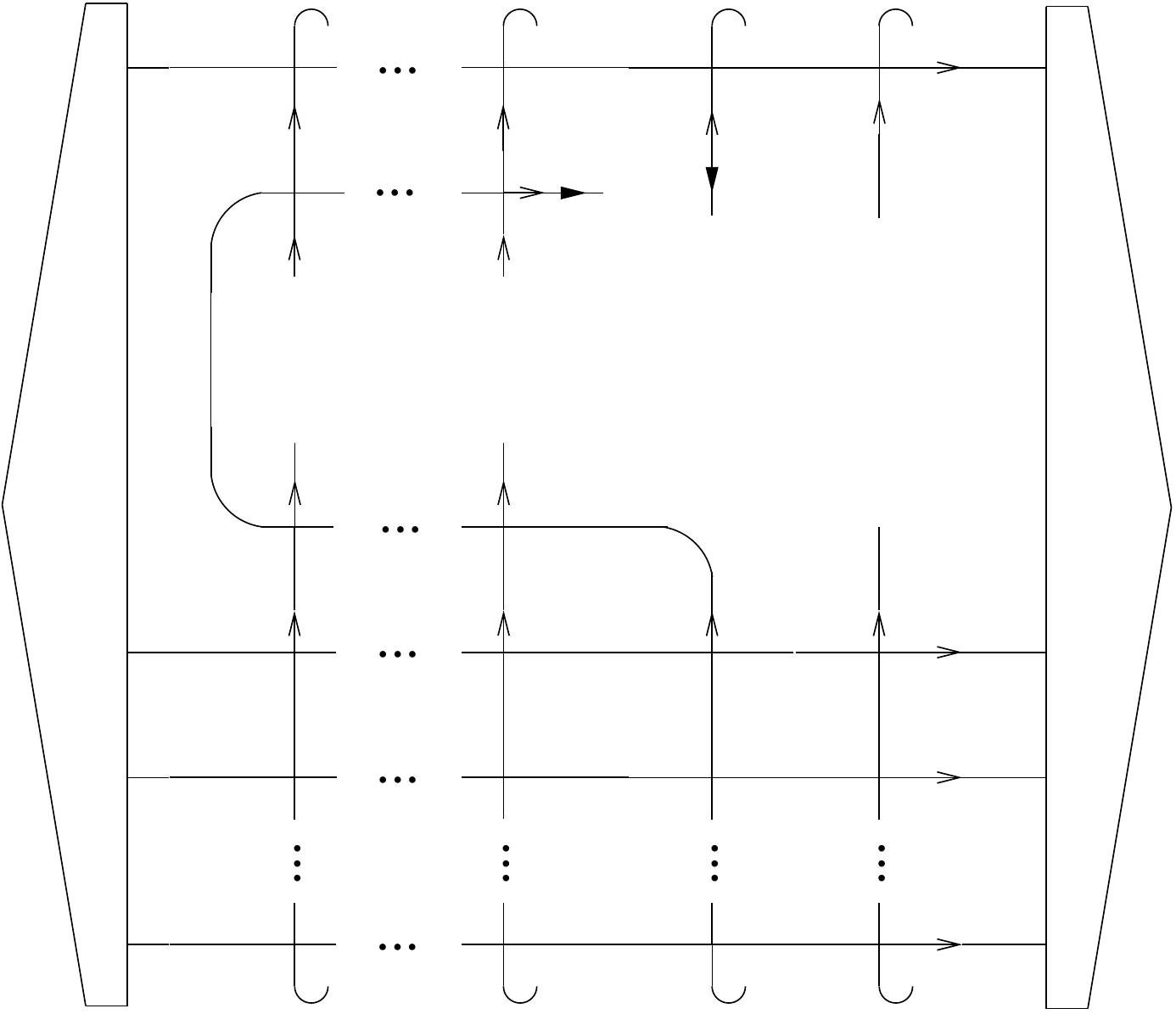}%
\end{picture}%
\setlength{\unitlength}{4144sp}%
\begingroup\makeatletter\ifx\SetFigFont\undefined%
\gdef\SetFigFont#1#2#3#4#5{%
  \reset@font\fontsize{#1}{#2pt}%
  \fontfamily{#3}\fontseries{#4}\fontshape{#5}%
  \selectfont}%
\fi\endgroup%
\begin{picture}(6324,5453)(439,-5382)
\put(676,-2746){\makebox(0,0)[lb]{\smash{{\SetFigFont{17}{20.4}{\familydefault}{\mddefault}{\updefault}{\color[rgb]{0,0,0}$\Phi_0$}%
}}}}
\put(2116,-3226){\makebox(0,0)[lb]{\smash{{\SetFigFont{17}{20.4}{\familydefault}{\mddefault}{\updefault}{\color[rgb]{0,0,0}$\lambda_1$}%
}}}}
\put(3256,-3226){\makebox(0,0)[lb]{\smash{{\SetFigFont{17}{20.4}{\familydefault}{\mddefault}{\updefault}{\color[rgb]{0,0,0}$\lambda_{n-1}$}%
}}}}
\put(4364,-3227){\makebox(0,0)[lb]{\smash{{\SetFigFont{17}{20.4}{\familydefault}{\mddefault}{\updefault}{\color[rgb]{0,0,0}$\la_n$}%
}}}}
\put(6256,-2761){\makebox(0,0)[lb]{\smash{{\SetFigFont{17}{20.4}{\familydefault}{\mddefault}{\updefault}{\color[rgb]{0,0,0}$\Phi_0$}%
}}}}
\put(5448,-211){\makebox(0,0)[lb]{\smash{{\SetFigFont{17}{20.4}{\familydefault}{\mddefault}{\updefault}{\color[rgb]{0,0,0}$\nu_N$}%
}}}}
\put(5463,-3735){\makebox(0,0)[lb]{\smash{{\SetFigFont{17}{20.4}{\familydefault}{\mddefault}{\updefault}{\color[rgb]{0,0,0}$\nu_1$}%
}}}}
\put(5462,-5282){\makebox(0,0)[lb]{\smash{{\SetFigFont{17}{20.4}{\familydefault}{\mddefault}{\updefault}{\color[rgb]{0,0,0}$\nu_{N-1}$}%
}}}}
\put(5463,-4380){\makebox(0,0)[lb]{\smash{{\SetFigFont{17}{20.4}{\familydefault}{\mddefault}{\updefault}{\color[rgb]{0,0,0}$\nu_2$}%
}}}}
\put(2478,-2789){\makebox(0,0)[lb]{\smash{{\SetFigFont{20}{24.0}{\rmdefault}{\mddefault}{\updefault}{\color[rgb]{0,0,0}...}%
}}}}
\put(2463,-3464){\makebox(0,0)[lb]{\smash{{\SetFigFont{20}{24.0}{\rmdefault}{\mddefault}{\updefault}{\color[rgb]{0,0,0}...}%
}}}}
\put(2448,-975){\makebox(0,0)[lb]{\smash{{\SetFigFont{20}{24.0}{\rmdefault}{\mddefault}{\updefault}{\color[rgb]{0,0,0}...}%
}}}}
\put(2463,-315){\makebox(0,0)[lb]{\smash{{\SetFigFont{20}{24.0}{\rmdefault}{\mddefault}{\updefault}{\color[rgb]{0,0,0}...}%
}}}}
\put(1936,-1621){\makebox(0,0)[lb]{\smash{{\SetFigFont{17}{20.4}{\familydefault}{\mddefault}{\updefault}{\color[rgb]{0,0,0}$\sigma_1$}%
}}}}
\put(3046,-1636){\makebox(0,0)[lb]{\smash{{\SetFigFont{17}{20.4}{\familydefault}{\mddefault}{\updefault}{\color[rgb]{0,0,0}$\sigma_{n-1}$}%
}}}}
\put(1951,-2251){\makebox(0,0)[lb]{\smash{{\SetFigFont{17}{20.4}{\familydefault}{\mddefault}{\updefault}{\color[rgb]{0,0,0}$\mu_1$}%
}}}}
\put(3031,-2266){\makebox(0,0)[lb]{\smash{{\SetFigFont{17}{20.4}{\familydefault}{\mddefault}{\updefault}{\color[rgb]{0,0,0}$\mu_{n-1}$}%
}}}}
\put(2463,-4139){\makebox(0,0)[lb]{\smash{{\SetFigFont{20}{24.0}{\rmdefault}{\mddefault}{\updefault}{\color[rgb]{0,0,0}...}%
}}}}
\put(2462,-5040){\makebox(0,0)[lb]{\smash{{\SetFigFont{20}{24.0}{\rmdefault}{\mddefault}{\updefault}{\color[rgb]{0,0,0}...}%
}}}}
\put(2027,-4456){\rotatebox{270.0}{\makebox(0,0)[lb]{\smash{{\SetFigFont{20}{24.0}{\rmdefault}{\mddefault}{\updefault}{\color[rgb]{0,0,0}...}%
}}}}}
\put(3152,-4456){\rotatebox{270.0}{\makebox(0,0)[lb]{\smash{{\SetFigFont{20}{24.0}{\rmdefault}{\mddefault}{\updefault}{\color[rgb]{0,0,0}...}%
}}}}}
\put(4277,-4456){\rotatebox{270.0}{\makebox(0,0)[lb]{\smash{{\SetFigFont{20}{24.0}{\rmdefault}{\mddefault}{\updefault}{\color[rgb]{0,0,0}...}%
}}}}}
\put(2133,-720){\makebox(0,0)[lb]{\smash{{\SetFigFont{17}{20.4}{\familydefault}{\mddefault}{\updefault}{\color[rgb]{0,0,0}$\lambda_1$}%
}}}}
\put(3258,-705){\makebox(0,0)[lb]{\smash{{\SetFigFont{17}{20.4}{\familydefault}{\mddefault}{\updefault}{\color[rgb]{0,0,0}$\lambda_{n-1}$}%
}}}}
\put(4381,-706){\makebox(0,0)[lb]{\smash{{\SetFigFont{17}{20.4}{\familydefault}{\mddefault}{\updefault}{\color[rgb]{0,0,0}$\la_n$}%
}}}}
\put(4171,-1306){\makebox(0,0)[lb]{\smash{{\SetFigFont{17}{20.4}{\familydefault}{\mddefault}{\updefault}{\color[rgb]{0,0,0}$\mu_n$}%
}}}}
\put(3736,-1066){\makebox(0,0)[lb]{\smash{{\SetFigFont{17}{20.4}{\familydefault}{\mddefault}{\updefault}{\color[rgb]{0,0,0}$\sigma_n$}%
}}}}
\put(5177,-4456){\rotatebox{270.0}{\makebox(0,0)[lb]{\smash{{\SetFigFont{20}{24.0}{\rmdefault}{\mddefault}{\updefault}{\color[rgb]{0,0,0}...}%
}}}}}
\put(5296,-3271){\makebox(0,0)[lb]{\smash{{\SetFigFont{17}{20.4}{\familydefault}{\mddefault}{\updefault}{\color[rgb]{0,0,0}$\la_n-1$}%
}}}}
\put(5266,-706){\makebox(0,0)[lb]{\smash{{\SetFigFont{17}{20.4}{\familydefault}{\mddefault}{\updefault}{\color[rgb]{0,0,0}$\la_n-1$}%
}}}}
\put(5026,-2701){\makebox(0,0)[lb]{\smash{{\SetFigFont{17}{20.4}{\familydefault}{\mddefault}{\updefault}{\color[rgb]{0,0,0}$\mu_{n+1}$}%
}}}}
\put(5071,-1306){\makebox(0,0)[lb]{\smash{{\SetFigFont{17}{20.4}{\familydefault}{\mddefault}{\updefault}{\color[rgb]{0,0,0}$\sigma_{n+1}$}%
}}}}
\end{picture}%

%% file: A_plus2.tex
\begin{picture}(0,0)%
\includegraphics{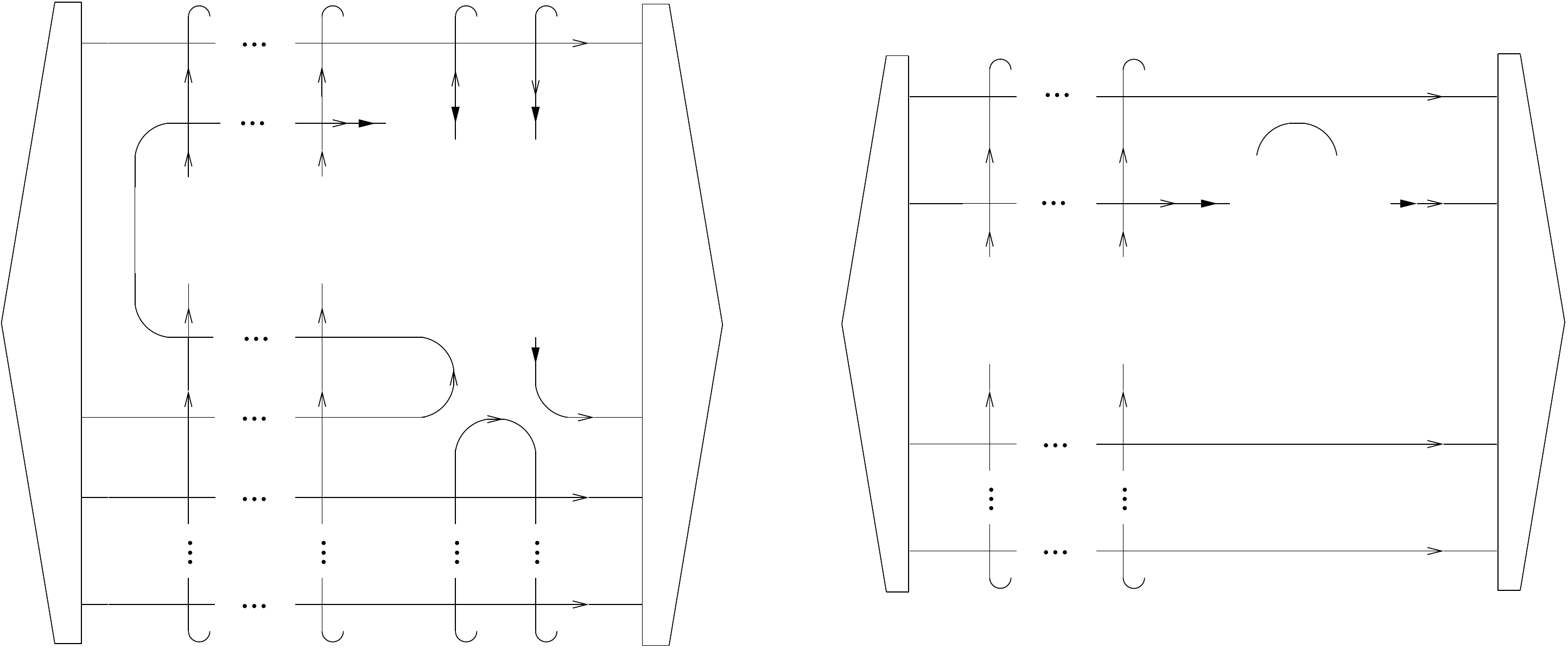}%
\end{picture}%
\setlength{\unitlength}{4144sp}%
\begingroup\makeatletter\ifx\SetFigFont\undefined%
\gdef\SetFigFont#1#2#3#4#5{%
  \reset@font\fontsize{#1}{#2pt}%
  \fontfamily{#3}\fontseries{#4}\fontshape{#5}%
  \selectfont}%
\fi\endgroup%
\begin{picture}(13194,5453)(439,-11457)
\put(676,-8821){\makebox(0,0)[lb]{\smash{{\SetFigFont{17}{20.4}{\familydefault}{\mddefault}{\updefault}{\color[rgb]{0,0,0}$\Phi_0$}%
}}}}
\put(5056,-6811){\makebox(0,0)[lb]{\smash{{\SetFigFont{17}{20.4}{\familydefault}{\mddefault}{\updefault}{\color[rgb]{0,0,0}$\nu_1$}%
}}}}
\put(4171,-7381){\makebox(0,0)[lb]{\smash{{\SetFigFont{17}{20.4}{\familydefault}{\mddefault}{\updefault}{\color[rgb]{0,0,0}$\mu_n$}%
}}}}
\put(4396,-6796){\makebox(0,0)[lb]{\smash{{\SetFigFont{17}{20.4}{\familydefault}{\mddefault}{\updefault}{\color[rgb]{0,0,0}$\nu_1$}%
}}}}
\put(2101,-9331){\makebox(0,0)[lb]{\smash{{\SetFigFont{17}{20.4}{\familydefault}{\mddefault}{\updefault}{\color[rgb]{0,0,0}$\lambda_1$}%
}}}}
\put(3226,-9316){\makebox(0,0)[lb]{\smash{{\SetFigFont{17}{20.4}{\familydefault}{\mddefault}{\updefault}{\color[rgb]{0,0,0}$\lambda_{n-1}$}%
}}}}
\put(6031,-8836){\makebox(0,0)[lb]{\smash{{\SetFigFont{17}{20.4}{\familydefault}{\mddefault}{\updefault}{\color[rgb]{0,0,0}$\Phi_0$}%
}}}}
\put(12437,-10907){\makebox(0,0)[lb]{\smash{{\SetFigFont{17}{20.4}{\familydefault}{\mddefault}{\updefault}{\color[rgb]{0,0,0}$\nu_{N-1}$}%
}}}}
\put(12438,-10005){\makebox(0,0)[lb]{\smash{{\SetFigFont{17}{20.4}{\familydefault}{\mddefault}{\updefault}{\color[rgb]{0,0,0}$\nu_2$}%
}}}}
\put(12423,-6736){\makebox(0,0)[lb]{\smash{{\SetFigFont{17}{20.4}{\familydefault}{\mddefault}{\updefault}{\color[rgb]{0,0,0}$\nu_N$}%
}}}}
\put(13216,-8896){\makebox(0,0)[lb]{\smash{{\SetFigFont{17}{20.4}{\familydefault}{\mddefault}{\updefault}{\color[rgb]{0,0,0}$\Phi_0$}%
}}}}
\put(8851,-7306){\makebox(0,0)[lb]{\smash{{\SetFigFont{17}{20.4}{\familydefault}{\mddefault}{\updefault}{\color[rgb]{0,0,0}$\lambda_1$}%
}}}}
\put(9976,-7291){\makebox(0,0)[lb]{\smash{{\SetFigFont{17}{20.4}{\familydefault}{\mddefault}{\updefault}{\color[rgb]{0,0,0}$\lambda_{n-1}$}%
}}}}
\put(7711,-8881){\makebox(0,0)[lb]{\smash{{\SetFigFont{17}{20.4}{\familydefault}{\mddefault}{\updefault}{\color[rgb]{0,0,0}$\Phi_0$}%
}}}}
\put(8851,-9556){\makebox(0,0)[lb]{\smash{{\SetFigFont{17}{20.4}{\familydefault}{\mddefault}{\updefault}{\color[rgb]{0,0,0}$\lambda_1$}%
}}}}
\put(9976,-9541){\makebox(0,0)[lb]{\smash{{\SetFigFont{17}{20.4}{\familydefault}{\mddefault}{\updefault}{\color[rgb]{0,0,0}$\lambda_{n-1}$}%
}}}}
\put(11581,-7770){\makebox(0,0)[lb]{\smash{{\SetFigFont{17}{20.4}{\familydefault}{\mddefault}{\updefault}{\color[rgb]{0,0,0}$\mu_{n+1}$}%
}}}}
\put(9211,-9764){\makebox(0,0)[lb]{\smash{{\SetFigFont{20}{24.0}{\rmdefault}{\mddefault}{\updefault}{\color[rgb]{0,0,0}...}%
}}}}
\put(9211,-10664){\makebox(0,0)[lb]{\smash{{\SetFigFont{20}{24.0}{\rmdefault}{\mddefault}{\updefault}{\color[rgb]{0,0,0}...}%
}}}}
\put(8776,-10080){\rotatebox{270.0}{\makebox(0,0)[lb]{\smash{{\SetFigFont{20}{24.0}{\rmdefault}{\mddefault}{\updefault}{\color[rgb]{0,0,0}...}%
}}}}}
\put(9196,-7725){\makebox(0,0)[lb]{\smash{{\SetFigFont{20}{24.0}{\rmdefault}{\mddefault}{\updefault}{\color[rgb]{0,0,0}...}%
}}}}
\put(9226,-6811){\makebox(0,0)[lb]{\smash{{\SetFigFont{20}{24.0}{\rmdefault}{\mddefault}{\updefault}{\color[rgb]{0,0,0}...}%
}}}}
\put(9766,-9016){\makebox(0,0)[lb]{\smash{{\SetFigFont{17}{20.4}{\familydefault}{\mddefault}{\updefault}{\color[rgb]{0,0,0}$\mu_{n-1}$}%
}}}}
\put(8656,-9016){\makebox(0,0)[lb]{\smash{{\SetFigFont{17}{20.4}{\familydefault}{\mddefault}{\updefault}{\color[rgb]{0,0,0}$\mu_1$}%
}}}}
\put(8671,-8371){\makebox(0,0)[lb]{\smash{{\SetFigFont{17}{20.4}{\familydefault}{\mddefault}{\updefault}{\color[rgb]{0,0,0}$\sigma_1$}%
}}}}
\put(9781,-8386){\makebox(0,0)[lb]{\smash{{\SetFigFont{17}{20.4}{\familydefault}{\mddefault}{\updefault}{\color[rgb]{0,0,0}$\sigma_{n-1}$}%
}}}}
\put(10141,-7995){\makebox(0,0)[lb]{\smash{{\SetFigFont{17}{20.4}{\familydefault}{\mddefault}{\updefault}{\color[rgb]{0,0,0}$\nu_1$}%
}}}}
\put(12423,-7995){\makebox(0,0)[lb]{\smash{{\SetFigFont{17}{20.4}{\familydefault}{\mddefault}{\updefault}{\color[rgb]{0,0,0}$\nu_1$}%
}}}}
\put(11575,-7471){\makebox(0,0)[lb]{\smash{{\SetFigFont{17}{20.4}{\familydefault}{\mddefault}{\updefault}{\color[rgb]{0,0,0}$\sigma_{n+1}$}%
}}}}
\put(10906,-7471){\makebox(0,0)[lb]{\smash{{\SetFigFont{17}{20.4}{\familydefault}{\mddefault}{\updefault}{\color[rgb]{0,0,0}$\mu_n$}%
}}}}
\put(10912,-7771){\makebox(0,0)[lb]{\smash{{\SetFigFont{17}{20.4}{\familydefault}{\mddefault}{\updefault}{\color[rgb]{0,0,0}$\sigma_n$}%
}}}}
\put(9901,-10080){\rotatebox{270.0}{\makebox(0,0)[lb]{\smash{{\SetFigFont{20}{24.0}{\rmdefault}{\mddefault}{\updefault}{\color[rgb]{0,0,0}...}%
}}}}}
\put(2478,-8864){\makebox(0,0)[lb]{\smash{{\SetFigFont{20}{24.0}{\rmdefault}{\mddefault}{\updefault}{\color[rgb]{0,0,0}...}%
}}}}
\put(2463,-9539){\makebox(0,0)[lb]{\smash{{\SetFigFont{20}{24.0}{\rmdefault}{\mddefault}{\updefault}{\color[rgb]{0,0,0}...}%
}}}}
\put(2448,-7050){\makebox(0,0)[lb]{\smash{{\SetFigFont{20}{24.0}{\rmdefault}{\mddefault}{\updefault}{\color[rgb]{0,0,0}...}%
}}}}
\put(2463,-6390){\makebox(0,0)[lb]{\smash{{\SetFigFont{20}{24.0}{\rmdefault}{\mddefault}{\updefault}{\color[rgb]{0,0,0}...}%
}}}}
\put(1936,-7696){\makebox(0,0)[lb]{\smash{{\SetFigFont{17}{20.4}{\familydefault}{\mddefault}{\updefault}{\color[rgb]{0,0,0}$\sigma_1$}%
}}}}
\put(3046,-7711){\makebox(0,0)[lb]{\smash{{\SetFigFont{17}{20.4}{\familydefault}{\mddefault}{\updefault}{\color[rgb]{0,0,0}$\sigma_{n-1}$}%
}}}}
\put(1951,-8326){\makebox(0,0)[lb]{\smash{{\SetFigFont{17}{20.4}{\familydefault}{\mddefault}{\updefault}{\color[rgb]{0,0,0}$\mu_1$}%
}}}}
\put(3031,-8341){\makebox(0,0)[lb]{\smash{{\SetFigFont{17}{20.4}{\familydefault}{\mddefault}{\updefault}{\color[rgb]{0,0,0}$\mu_{n-1}$}%
}}}}
\put(2463,-10214){\makebox(0,0)[lb]{\smash{{\SetFigFont{20}{24.0}{\rmdefault}{\mddefault}{\updefault}{\color[rgb]{0,0,0}...}%
}}}}
\put(2462,-11115){\makebox(0,0)[lb]{\smash{{\SetFigFont{20}{24.0}{\rmdefault}{\mddefault}{\updefault}{\color[rgb]{0,0,0}...}%
}}}}
\put(2027,-10531){\rotatebox{270.0}{\makebox(0,0)[lb]{\smash{{\SetFigFont{20}{24.0}{\rmdefault}{\mddefault}{\updefault}{\color[rgb]{0,0,0}...}%
}}}}}
\put(3152,-10531){\rotatebox{270.0}{\makebox(0,0)[lb]{\smash{{\SetFigFont{20}{24.0}{\rmdefault}{\mddefault}{\updefault}{\color[rgb]{0,0,0}...}%
}}}}}
\put(2133,-6795){\makebox(0,0)[lb]{\smash{{\SetFigFont{17}{20.4}{\familydefault}{\mddefault}{\updefault}{\color[rgb]{0,0,0}$\lambda_1$}%
}}}}
\put(3258,-6780){\makebox(0,0)[lb]{\smash{{\SetFigFont{17}{20.4}{\familydefault}{\mddefault}{\updefault}{\color[rgb]{0,0,0}$\lambda_{n-1}$}%
}}}}
\put(3736,-7141){\makebox(0,0)[lb]{\smash{{\SetFigFont{17}{20.4}{\familydefault}{\mddefault}{\updefault}{\color[rgb]{0,0,0}$\sigma_n$}%
}}}}
\put(4846,-7381){\makebox(0,0)[lb]{\smash{{\SetFigFont{17}{20.4}{\familydefault}{\mddefault}{\updefault}{\color[rgb]{0,0,0}$\sigma_{n+1}$}%
}}}}
\put(4276,-10531){\rotatebox{270.0}{\makebox(0,0)[lb]{\smash{{\SetFigFont{20}{24.0}{\rmdefault}{\mddefault}{\updefault}{\color[rgb]{0,0,0}...}%
}}}}}
\put(4951,-10531){\rotatebox{270.0}{\makebox(0,0)[lb]{\smash{{\SetFigFont{20}{24.0}{\rmdefault}{\mddefault}{\updefault}{\color[rgb]{0,0,0}...}%
}}}}}
\put(4801,-8776){\makebox(0,0)[lb]{\smash{{\SetFigFont{17}{20.4}{\familydefault}{\mddefault}{\updefault}{\color[rgb]{0,0,0}$\mu_{n+1}$}%
}}}}
\put(4381,-9271){\makebox(0,0)[lb]{\smash{{\SetFigFont{17}{20.4}{\familydefault}{\mddefault}{\updefault}{\color[rgb]{0,0,0}$\nu_1$}%
}}}}
\put(4531,-9811){\makebox(0,0)[lb]{\smash{{\SetFigFont{17}{20.4}{\familydefault}{\mddefault}{\updefault}{\color[rgb]{0,0,0}$\nu_1$}%
}}}}
\put(5223,-6286){\makebox(0,0)[lb]{\smash{{\SetFigFont{17}{20.4}{\familydefault}{\mddefault}{\updefault}{\color[rgb]{0,0,0}$\nu_N$}%
}}}}
\put(5238,-9810){\makebox(0,0)[lb]{\smash{{\SetFigFont{17}{20.4}{\familydefault}{\mddefault}{\updefault}{\color[rgb]{0,0,0}$\nu_1$}%
}}}}
\put(5237,-11357){\makebox(0,0)[lb]{\smash{{\SetFigFont{17}{20.4}{\familydefault}{\mddefault}{\updefault}{\color[rgb]{0,0,0}$\nu_{N-1}$}%
}}}}
\put(5238,-10455){\makebox(0,0)[lb]{\smash{{\SetFigFont{17}{20.4}{\familydefault}{\mddefault}{\updefault}{\color[rgb]{0,0,0}$\nu_2$}%
}}}}
\put(6901,-8866){\makebox(0,0)[lb]{\smash{{\SetFigFont{17}{20.4}{\familydefault}{\mddefault}{\updefault}{\color[rgb]{0,0,0}$\propto$}%
}}}}
\end{picture}%

%% file: FE_mod.tex
\begin{picture}(0,0)%
\includegraphics{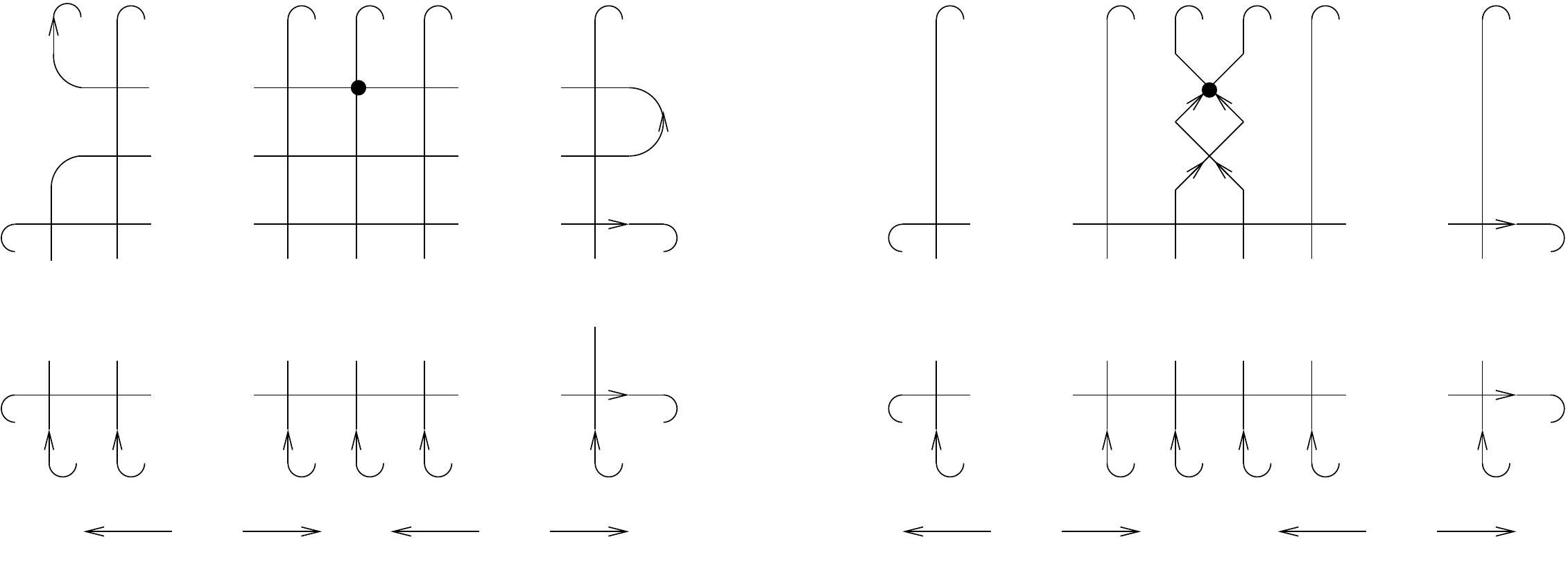}%
\end{picture}%
\setlength{\unitlength}{4144sp}%
\begingroup\makeatletter\ifx\SetFigFont\undefined%
\gdef\SetFigFont#1#2#3#4#5{%
  \reset@font\fontsize{#1}{#2pt}%
  \fontfamily{#3}\fontseries{#4}\fontshape{#5}%
  \selectfont}%
\fi\endgroup%
\begin{picture}(10323,3700)(577,-10397)
\put(8416,-9661){\makebox(0,0)[lb]{\smash{{\SetFigFont{17}{20.4}{\familydefault}{\mddefault}{\updefault}{\color[rgb]{0,0,0}$\mu$}%
}}}}
\put(8866,-9661){\makebox(0,0)[lb]{\smash{{\SetFigFont{17}{20.4}{\familydefault}{\mddefault}{\updefault}{\color[rgb]{0,0,0}$\lambda$}%
}}}}
\put(1846,-10306){\makebox(0,0)[lb]{\smash{{\SetFigFont{17}{20.4}{\familydefault}{\mddefault}{\updefault}{\color[rgb]{0,0,0}$k$}%
}}}}
\put(3871,-10306){\makebox(0,0)[lb]{\smash{{\SetFigFont{17}{20.4}{\familydefault}{\mddefault}{\updefault}{\color[rgb]{0,0,0}$k$}%
}}}}
\put(7246,-10306){\makebox(0,0)[lb]{\smash{{\SetFigFont{17}{20.4}{\familydefault}{\mddefault}{\updefault}{\color[rgb]{0,0,0}$k$}%
}}}}
\put(9721,-10306){\makebox(0,0)[lb]{\smash{{\SetFigFont{17}{20.4}{\familydefault}{\mddefault}{\updefault}{\color[rgb]{0,0,0}$k$}%
}}}}
\put(1771,-7291){\makebox(0,0)[lb]{\smash{{\SetFigFont{12}{14.4}{\rmdefault}{\mddefault}{\updefault}{\color[rgb]{0,0,0}...}%
}}}}
\put(3766,-7276){\makebox(0,0)[lb]{\smash{{\SetFigFont{12}{14.4}{\rmdefault}{\mddefault}{\updefault}{\color[rgb]{0,0,0}...}%
}}}}
\put(1741,-7726){\makebox(0,0)[lb]{\smash{{\SetFigFont{12}{14.4}{\rmdefault}{\mddefault}{\updefault}{\color[rgb]{0,0,0}...}%
}}}}
\put(3766,-7726){\makebox(0,0)[lb]{\smash{{\SetFigFont{12}{14.4}{\rmdefault}{\mddefault}{\updefault}{\color[rgb]{0,0,0}...}%
}}}}
\put(1741,-9301){\makebox(0,0)[lb]{\smash{{\SetFigFont{12}{14.4}{\rmdefault}{\mddefault}{\updefault}{\color[rgb]{0,0,0}...}%
}}}}
\put(3766,-9301){\makebox(0,0)[lb]{\smash{{\SetFigFont{12}{14.4}{\rmdefault}{\mddefault}{\updefault}{\color[rgb]{0,0,0}...}%
}}}}
\put(1741,-8176){\makebox(0,0)[lb]{\smash{{\SetFigFont{12}{14.4}{\rmdefault}{\mddefault}{\updefault}{\color[rgb]{0,0,0}...}%
}}}}
\put(3766,-8176){\makebox(0,0)[lb]{\smash{{\SetFigFont{12}{14.4}{\rmdefault}{\mddefault}{\updefault}{\color[rgb]{0,0,0}...}%
}}}}
\put(886,-8806){\makebox(0,0)[lb]{\smash{{\SetFigFont{12}{14.4}{\familydefault}{\mddefault}{\updefault}{\color[rgb]{0,0,0}$\vdots$}%
}}}}
\put(2461,-8791){\makebox(0,0)[lb]{\smash{{\SetFigFont{12}{14.4}{\familydefault}{\mddefault}{\updefault}{\color[rgb]{0,0,0}$\vdots$}%
}}}}
\put(2926,-8791){\makebox(0,0)[lb]{\smash{{\SetFigFont{12}{14.4}{\familydefault}{\mddefault}{\updefault}{\color[rgb]{0,0,0}$\vdots$}%
}}}}
\put(3376,-8791){\makebox(0,0)[lb]{\smash{{\SetFigFont{12}{14.4}{\familydefault}{\mddefault}{\updefault}{\color[rgb]{0,0,0}$\vdots$}%
}}}}
\put(4501,-8776){\makebox(0,0)[lb]{\smash{{\SetFigFont{12}{14.4}{\familydefault}{\mddefault}{\updefault}{\color[rgb]{0,0,0}$\vdots$}%
}}}}
\put(1336,-8791){\makebox(0,0)[lb]{\smash{{\SetFigFont{12}{14.4}{\familydefault}{\mddefault}{\updefault}{\color[rgb]{0,0,0}$\vdots$}%
}}}}
\put(4591,-8086){\makebox(0,0)[lb]{\smash{{\SetFigFont{17}{20.4}{\familydefault}{\mddefault}{\updefault}{\color[rgb]{0,0,0}$\nu_1$}%
}}}}
\put(4561,-9211){\makebox(0,0)[lb]{\smash{{\SetFigFont{17}{20.4}{\familydefault}{\mddefault}{\updefault}{\color[rgb]{0,0,0}$\nu_N$}%
}}}}
\put(5011,-7591){\makebox(0,0)[lb]{\smash{{\SetFigFont{17}{20.4}{\familydefault}{\mddefault}{\updefault}{\color[rgb]{0,0,0}$\mu$}%
}}}}
\put(961,-9616){\makebox(0,0)[lb]{\smash{{\SetFigFont{17}{20.4}{\familydefault}{\mddefault}{\updefault}{\color[rgb]{0,0,0}$\mu$}%
}}}}
\put(1006,-7081){\makebox(0,0)[lb]{\smash{{\SetFigFont{17}{20.4}{\familydefault}{\mddefault}{\updefault}{\color[rgb]{0,0,0}$\mu$}%
}}}}
\put(9616,-9301){\makebox(0,0)[lb]{\smash{{\SetFigFont{12}{14.4}{\rmdefault}{\mddefault}{\updefault}{\color[rgb]{0,0,0}...}%
}}}}
\put(9616,-8176){\makebox(0,0)[lb]{\smash{{\SetFigFont{12}{14.4}{\rmdefault}{\mddefault}{\updefault}{\color[rgb]{0,0,0}...}%
}}}}
\put(8311,-8791){\makebox(0,0)[lb]{\smash{{\SetFigFont{12}{14.4}{\familydefault}{\mddefault}{\updefault}{\color[rgb]{0,0,0}$\vdots$}%
}}}}
\put(8776,-8791){\makebox(0,0)[lb]{\smash{{\SetFigFont{12}{14.4}{\familydefault}{\mddefault}{\updefault}{\color[rgb]{0,0,0}$\vdots$}%
}}}}
\put(9226,-8791){\makebox(0,0)[lb]{\smash{{\SetFigFont{12}{14.4}{\familydefault}{\mddefault}{\updefault}{\color[rgb]{0,0,0}$\vdots$}%
}}}}
\put(10351,-8776){\makebox(0,0)[lb]{\smash{{\SetFigFont{12}{14.4}{\familydefault}{\mddefault}{\updefault}{\color[rgb]{0,0,0}$\vdots$}%
}}}}
\put(7861,-8806){\makebox(0,0)[lb]{\smash{{\SetFigFont{12}{14.4}{\familydefault}{\mddefault}{\updefault}{\color[rgb]{0,0,0}$\vdots$}%
}}}}
\put(7141,-9301){\makebox(0,0)[lb]{\smash{{\SetFigFont{12}{14.4}{\rmdefault}{\mddefault}{\updefault}{\color[rgb]{0,0,0}...}%
}}}}
\put(7141,-8176){\makebox(0,0)[lb]{\smash{{\SetFigFont{12}{14.4}{\rmdefault}{\mddefault}{\updefault}{\color[rgb]{0,0,0}...}%
}}}}
\put(6736,-8791){\makebox(0,0)[lb]{\smash{{\SetFigFont{12}{14.4}{\familydefault}{\mddefault}{\updefault}{\color[rgb]{0,0,0}$\vdots$}%
}}}}
\put(10411,-8056){\makebox(0,0)[lb]{\smash{{\SetFigFont{17}{20.4}{\familydefault}{\mddefault}{\updefault}{\color[rgb]{0,0,0}$\nu_1$}%
}}}}
\put(10411,-9181){\makebox(0,0)[lb]{\smash{{\SetFigFont{17}{20.4}{\familydefault}{\mddefault}{\updefault}{\color[rgb]{0,0,0}$\nu_N$}%
}}}}
\put(5626,-8611){\makebox(0,0)[lb]{\smash{{\SetFigFont{17}{20.4}{\familydefault}{\mddefault}{\updefault}{\color[rgb]{0,0,0}$\propto$}%
}}}}
\put(2986,-9661){\makebox(0,0)[lb]{\smash{{\SetFigFont{17}{20.4}{\familydefault}{\mddefault}{\updefault}{\color[rgb]{0,0,0}$\lambda$}%
}}}}
\end{picture}%

%% file: R_dot.tex
\begin{picture}(0,0)%
\includegraphics{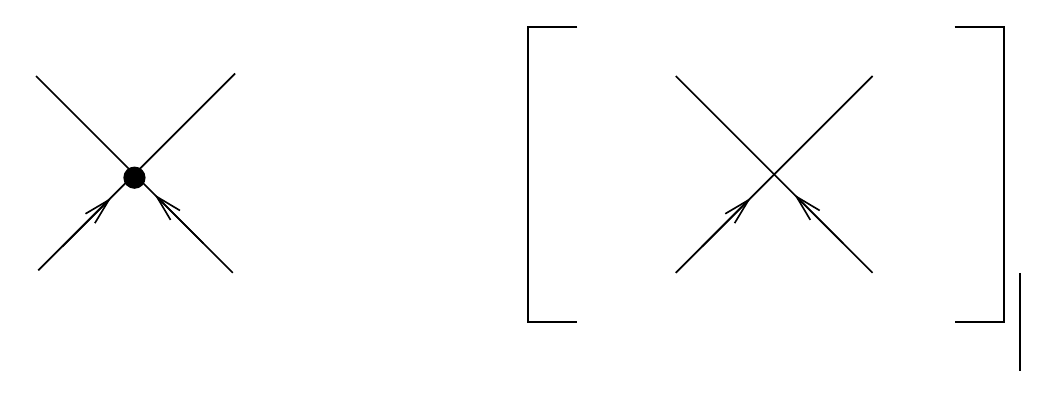}%
\end{picture}%
\setlength{\unitlength}{4144sp}%
\begingroup\makeatletter\ifx\SetFigFont\undefined%
\gdef\SetFigFont#1#2#3#4#5{%
  \reset@font\fontsize{#1}{#2pt}%
  \fontfamily{#3}\fontseries{#4}\fontshape{#5}%
  \selectfont}%
\fi\endgroup%
\begin{picture}(4785,1846)(1636,-2471)
\put(1681,-2071){\makebox(0,0)[lb]{\smash{{\SetFigFont{17}{20.4}{\familydefault}{\mddefault}{\updefault}{\color[rgb]{0,0,0}$\sigma_1$}%
}}}}
\put(1831,-1636){\makebox(0,0)[lb]{\smash{{\SetFigFont{17}{20.4}{\familydefault}{\mddefault}{\updefault}{\color[rgb]{0,0,0}$\la$}%
}}}}
\put(2551,-1651){\makebox(0,0)[lb]{\smash{{\SetFigFont{17}{20.4}{\familydefault}{\mddefault}{\updefault}{\color[rgb]{0,0,0}$\mu$}%
}}}}
\put(2656,-856){\makebox(0,0)[lb]{\smash{{\SetFigFont{17}{20.4}{\familydefault}{\mddefault}{\updefault}{\color[rgb]{0,0,0}$\mu_2$}%
}}}}
\put(1651,-886){\makebox(0,0)[lb]{\smash{{\SetFigFont{17}{20.4}{\familydefault}{\mddefault}{\updefault}{\color[rgb]{0,0,0}$\mu_1$}%
}}}}
\put(6406,-2371){\makebox(0,0)[lb]{\smash{{\SetFigFont{17}{20.4}{\familydefault}{\mddefault}{\updefault}{\color[rgb]{0,0,0}$\varepsilon=0$}%
}}}}
\put(4336,-1486){\makebox(0,0)[lb]{\smash{{\SetFigFont{17}{20.4}{\familydefault}{\mddefault}{\updefault}{\color[rgb]{0,0,0}$\frac{d}{d\varepsilon}$}%
}}}}
\put(3286,-1486){\makebox(0,0)[lb]{\smash{{\SetFigFont{17}{20.4}{\familydefault}{\mddefault}{\updefault}{\color[rgb]{0,0,0}$:=$}%
}}}}
\put(4591,-2086){\makebox(0,0)[lb]{\smash{{\SetFigFont{17}{20.4}{\familydefault}{\mddefault}{\updefault}{\color[rgb]{0,0,0}$\sigma_1$}%
}}}}
\put(4756,-1636){\makebox(0,0)[lb]{\smash{{\SetFigFont{17}{20.4}{\familydefault}{\mddefault}{\updefault}{\color[rgb]{0,0,0}$\la$}%
}}}}
\put(4576,-871){\makebox(0,0)[lb]{\smash{{\SetFigFont{17}{20.4}{\familydefault}{\mddefault}{\updefault}{\color[rgb]{0,0,0}$\mu_1$}%
}}}}
\put(5551,-886){\makebox(0,0)[lb]{\smash{{\SetFigFont{17}{20.4}{\familydefault}{\mddefault}{\updefault}{\color[rgb]{0,0,0}$\mu_2$}%
}}}}
\put(5476,-1636){\makebox(0,0)[lb]{\smash{{\SetFigFont{17}{20.4}{\familydefault}{\mddefault}{\updefault}{\color[rgb]{0,0,0}$\mu+\varepsilon$}%
}}}}
\put(5566,-2086){\makebox(0,0)[lb]{\smash{{\SetFigFont{17}{20.4}{\familydefault}{\mddefault}{\updefault}{\color[rgb]{0,0,0}$\sigma_2$}%
}}}}
\put(2656,-2086){\makebox(0,0)[lb]{\smash{{\SetFigFont{17}{20.4}{\familydefault}{\mddefault}{\updefault}{\color[rgb]{0,0,0}$\sigma_2$}%
}}}}
\end{picture}%

%% file: A_alpha.tex
\begin{picture}(0,0)%
\includegraphics{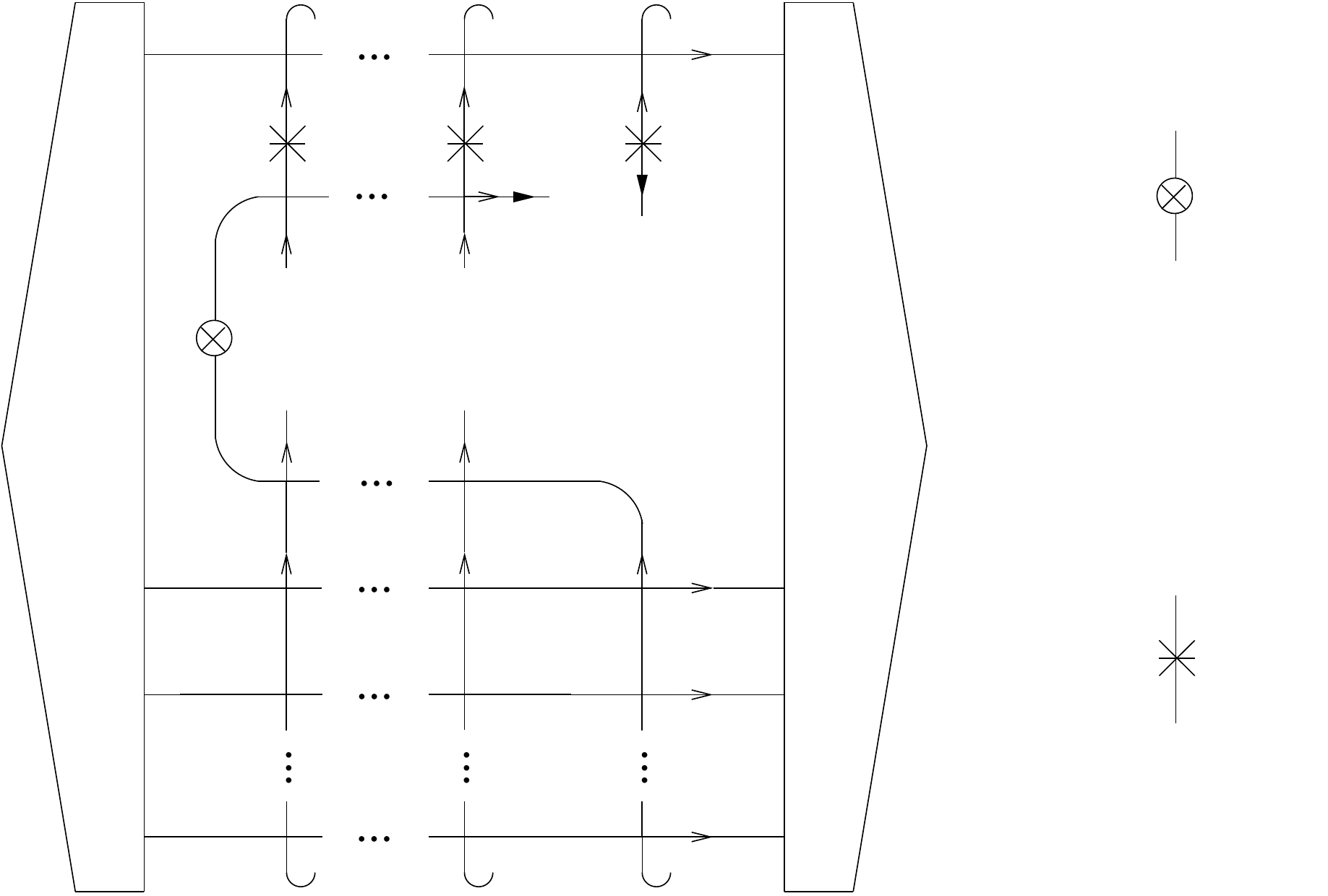}%
\end{picture}%
\setlength{\unitlength}{4144sp}%
\begingroup\makeatletter\ifx\SetFigFont\undefined%
\gdef\SetFigFont#1#2#3#4#5{%
  \reset@font\fontsize{#1}{#2pt}%
  \fontfamily{#3}\fontseries{#4}\fontshape{#5}%
  \selectfont}%
\fi\endgroup%
\begin{picture}(8397,5663)(664,-5382)
\put(2566,-3226){\makebox(0,0)[lb]{\smash{{\SetFigFont{17}{20.4}{\familydefault}{\mddefault}{\updefault}{\color[rgb]{0,0,0}$\lambda_1$}%
}}}}
\put(3706,-3226){\makebox(0,0)[lb]{\smash{{\SetFigFont{17}{20.4}{\familydefault}{\mddefault}{\updefault}{\color[rgb]{0,0,0}$\lambda_{n-1}$}%
}}}}
\put(4814,-3227){\makebox(0,0)[lb]{\smash{{\SetFigFont{17}{20.4}{\familydefault}{\mddefault}{\updefault}{\color[rgb]{0,0,0}$\la_n$}%
}}}}
\put(706,-2641){\makebox(0,0)[lb]{\smash{{\SetFigFont{17}{20.4}{\familydefault}{\mddefault}{\updefault}{\color[rgb]{0,0,0}$\Phi_0^{2\alpha\eta+\beta h}$}%
}}}}
\put(8011,-511){\makebox(0,0)[lb]{\smash{{\SetFigFont{17}{20.4}{\familydefault}{\mddefault}{\updefault}{\color[rgb]{0,0,0}$\mu$}%
}}}}
\put(8506,-1066){\makebox(0,0)[lb]{\smash{{\SetFigFont{17}{20.4}{\familydefault}{\mddefault}{\updefault}{\color[rgb]{0,0,0}$=$}%
}}}}
\put(7996,-1576){\makebox(0,0)[lb]{\smash{{\SetFigFont{17}{20.4}{\familydefault}{\mddefault}{\updefault}{\color[rgb]{0,0,0}$\sigma$}%
}}}}
\put(9046,-1051){\makebox(0,0)[lb]{\smash{{\SetFigFont{17}{20.4}{\familydefault}{\mddefault}{\updefault}{\color[rgb]{0,0,0}$\begin{pmatrix}q^{2\alpha}&0\\0&q^{-2\alpha}\end{pmatrix}_{\mu, \sigma}$}%
}}}}
\put(7996,-3451){\makebox(0,0)[lb]{\smash{{\SetFigFont{17}{20.4}{\familydefault}{\mddefault}{\updefault}{\color[rgb]{0,0,0}$\mu$}%
}}}}
\put(7996,-4486){\makebox(0,0)[lb]{\smash{{\SetFigFont{17}{20.4}{\familydefault}{\mddefault}{\updefault}{\color[rgb]{0,0,0}$\sigma$}%
}}}}
\put(8491,-3991){\makebox(0,0)[lb]{\smash{{\SetFigFont{17}{20.4}{\familydefault}{\mddefault}{\updefault}{\color[rgb]{0,0,0}$=$}%
}}}}
\put(8941,-3976){\makebox(0,0)[lb]{\smash{{\SetFigFont{17}{20.4}{\familydefault}{\mddefault}{\updefault}{\color[rgb]{0,0,0}$\begin{pmatrix}e^{\beta h}&0\\0&e^{-\beta h}\end{pmatrix}_{\mu, \sigma}$}%
}}}}
\put(2928,-2789){\makebox(0,0)[lb]{\smash{{\SetFigFont{20}{24.0}{\rmdefault}{\mddefault}{\updefault}{\color[rgb]{0,0,0}...}%
}}}}
\put(2913,-3464){\makebox(0,0)[lb]{\smash{{\SetFigFont{20}{24.0}{\rmdefault}{\mddefault}{\updefault}{\color[rgb]{0,0,0}...}%
}}}}
\put(2898,-975){\makebox(0,0)[lb]{\smash{{\SetFigFont{20}{24.0}{\rmdefault}{\mddefault}{\updefault}{\color[rgb]{0,0,0}...}%
}}}}
\put(2386,-1621){\makebox(0,0)[lb]{\smash{{\SetFigFont{17}{20.4}{\familydefault}{\mddefault}{\updefault}{\color[rgb]{0,0,0}$\sigma_1$}%
}}}}
\put(3496,-1636){\makebox(0,0)[lb]{\smash{{\SetFigFont{17}{20.4}{\familydefault}{\mddefault}{\updefault}{\color[rgb]{0,0,0}$\sigma_{n-1}$}%
}}}}
\put(2401,-2251){\makebox(0,0)[lb]{\smash{{\SetFigFont{17}{20.4}{\familydefault}{\mddefault}{\updefault}{\color[rgb]{0,0,0}$\mu_1$}%
}}}}
\put(3481,-2266){\makebox(0,0)[lb]{\smash{{\SetFigFont{17}{20.4}{\familydefault}{\mddefault}{\updefault}{\color[rgb]{0,0,0}$\mu_{n-1}$}%
}}}}
\put(2913,-4139){\makebox(0,0)[lb]{\smash{{\SetFigFont{20}{24.0}{\rmdefault}{\mddefault}{\updefault}{\color[rgb]{0,0,0}...}%
}}}}
\put(2912,-5040){\makebox(0,0)[lb]{\smash{{\SetFigFont{20}{24.0}{\rmdefault}{\mddefault}{\updefault}{\color[rgb]{0,0,0}...}%
}}}}
\put(2477,-4456){\rotatebox{270.0}{\makebox(0,0)[lb]{\smash{{\SetFigFont{20}{24.0}{\rmdefault}{\mddefault}{\updefault}{\color[rgb]{0,0,0}...}%
}}}}}
\put(3602,-4456){\rotatebox{270.0}{\makebox(0,0)[lb]{\smash{{\SetFigFont{20}{24.0}{\rmdefault}{\mddefault}{\updefault}{\color[rgb]{0,0,0}...}%
}}}}}
\put(4727,-4456){\rotatebox{270.0}{\makebox(0,0)[lb]{\smash{{\SetFigFont{20}{24.0}{\rmdefault}{\mddefault}{\updefault}{\color[rgb]{0,0,0}...}%
}}}}}
\put(5013,-3735){\makebox(0,0)[lb]{\smash{{\SetFigFont{17}{20.4}{\familydefault}{\mddefault}{\updefault}{\color[rgb]{0,0,0}$\nu_1$}%
}}}}
\put(5012,-5282){\makebox(0,0)[lb]{\smash{{\SetFigFont{17}{20.4}{\familydefault}{\mddefault}{\updefault}{\color[rgb]{0,0,0}$\nu_{N-1}$}%
}}}}
\put(5013,-4380){\makebox(0,0)[lb]{\smash{{\SetFigFont{17}{20.4}{\familydefault}{\mddefault}{\updefault}{\color[rgb]{0,0,0}$\nu_2$}%
}}}}
\put(4621,-1306){\makebox(0,0)[lb]{\smash{{\SetFigFont{17}{20.4}{\familydefault}{\mddefault}{\updefault}{\color[rgb]{0,0,0}$\mu_n$}%
}}}}
\put(4186,-1066){\makebox(0,0)[lb]{\smash{{\SetFigFont{17}{20.4}{\familydefault}{\mddefault}{\updefault}{\color[rgb]{0,0,0}$\sigma_n$}%
}}}}
\put(2913,-90){\makebox(0,0)[lb]{\smash{{\SetFigFont{20}{24.0}{\rmdefault}{\mddefault}{\updefault}{\color[rgb]{0,0,0}...}%
}}}}
\put(2583,-495){\makebox(0,0)[lb]{\smash{{\SetFigFont{17}{20.4}{\familydefault}{\mddefault}{\updefault}{\color[rgb]{0,0,0}$\lambda_1$}%
}}}}
\put(3708,-480){\makebox(0,0)[lb]{\smash{{\SetFigFont{17}{20.4}{\familydefault}{\mddefault}{\updefault}{\color[rgb]{0,0,0}$\lambda_{n-1}$}%
}}}}
\put(4831,-481){\makebox(0,0)[lb]{\smash{{\SetFigFont{17}{20.4}{\familydefault}{\mddefault}{\updefault}{\color[rgb]{0,0,0}$\la_n$}%
}}}}
\put(4998, 14){\makebox(0,0)[lb]{\smash{{\SetFigFont{17}{20.4}{\familydefault}{\mddefault}{\updefault}{\color[rgb]{0,0,0}$\nu_N$}%
}}}}
\put(5866,-2641){\makebox(0,0)[lb]{\smash{{\SetFigFont{17}{20.4}{\familydefault}{\mddefault}{\updefault}{\color[rgb]{0,0,0}$\Phi_0^{\beta h}$}%
}}}}
\end{picture}%